\documentclass[%
 reprint,
superscriptaddress,
preprintnumbers,
nofootinbib,
 amsmath,amssymb,
 aps,
]{revtex4-1}

\usepackage[normalem]{ulem}
\usepackage[colorlinks]{hyperref}
\usepackage{graphicx}
\usepackage{dcolumn}
\usepackage{bm}
\usepackage{color}
\usepackage{bbold}
\usepackage[dvipsnames]{xcolor}
\usepackage{placeins}
\usepackage{slashed}
\usepackage{multirow}
\usepackage{subfigure}
\usepackage{tabularx}
\usepackage{mathtools}
\usepackage{blindtext}
\usepackage{soul}
\usepackage{slashed}
\usepackage{amsmath}
\usepackage{orcidlink}

\hypersetup{
  colorlinks=true,
  linkcolor=red,
  citecolor=blue,
  urlcolor=blue
}

\definecolor{coral}{RGB}{255,127,80}
\definecolor{indigo}{RGB}{75,0,130}
\definecolor{red}{rgb}{0.9, 0,0}
\definecolor{cerulean}{rgb}{0., 0.62,0.9}
\definecolor{navy}{rgb}{0.05, 0.05,0.8}

\newcommand{\mZ}{m_{Z^\prime}}

\newcommand{\TeV}{{\rm TeV}}
\newcommand{\GeV}{{\rm GeV}}
\newcommand{\MeV}{{\rm MeV}}

\begin{document}
\preprint{CERN-TH-2023-134,TTP23-023, FR-PHENO-2023-06, P3H-23-042}
\title{Supernova Limits on Muonic Dark Forces}

\author{Claudio Andrea Manzari\,\orcidlink{0000-0001-8114-3078}}
\email{camanzari@lbl.gov}
\affiliation{Berkeley Center for Theoretical Physics, Department of Physics, University of California, Berkeley, CA 94720, USA}
\affiliation{Theoretical Physics Group, Lawrence Berkeley National Laboratory, Berkeley, CA 94720, USA}
\author{Jorge Martin Camalich\,\orcidlink{0000-0002-5673-4500}}
\email{jcamalich@iac.es}
\affiliation{Instituto de Astrof\'isica de Canarias,  C/ V\'ia L\'actea, s/n E38205 - La Laguna, Tenerife, Spain}
\affiliation{Universidad de La Laguna, Departamento de Astrof\'isica, La Laguna, Tenerife, Spain}
\affiliation{CERN, Theoretical Physics Department, CH-1211 Geneva 23, Switzerland}
\author{Jonas Spinner\,\orcidlink{0000-0003-4046-948X}}
\email{j.spinner@thphys.uni-heidelberg.de}
\affiliation{Institut f\"ur Theoretische Teilchenphysik, Karlsruhe Institute of Technology, Karlsruhe, Germany}
\affiliation{Institut f\"ur Theoretische Physik, Universit\"at Heidelberg, Germany}
\author{Robert Ziegler\,\orcidlink{0000-0003-2653-7327}}
\email{robert.ziegler@kit.edu}
\affiliation{Institut f\"ur Theoretische Teilchenphysik, Karlsruhe Institute of Technology, Karlsruhe, Germany}
\affiliation{Physikalisches Institut, Albert-Ludwigs-Universit\"at Freiburg, 79104 Freiburg, Germany}
\date{\today}

\begin{abstract}
Proto-neutron stars formed during core-collapse supernovae are hot and dense environments that contain a sizable population of muons. If these interact with new long-lived particles with masses up to roughly 100 MeV,  the latter can be produced and escape from the stellar plasma, causing an excessive energy loss constrained by  observations of SN~1987A. In this article we calculate the emission of light dark fermions that are coupled to leptons via a new massive vector boson, and determine the resulting constraints on the general parameter space. We apply these limits to the gauged $L_\mu-L_\tau$ model with dark fermions, and show that the SN~1987A constraints exclude a significant portion of the parameter space targeted by future experiments. We also extend our analysis to generic effective four-fermion operators that couple dark fermions to muons, electrons, or neutrinos. We find that SN~1987A cooling probes a new-physics scale up to $\sim7$ TeV, which is an order of magnitude larger than current bounds from laboratory experiments.
\end{abstract}

\maketitle

\section{Introduction}
\label{sec:intro}

Understanding the fundamental nature of Dark Matter (DM), which comprises $\sim84\%$ of the matter of the Universe~\cite{Planck:2018vyg}, has become one of the most pressing problems in contemporary physics (see Ref.~\cite{Bertone:2010zza} for a review). A wide class of theoretical models describe DM as new light (sub-GeV) particles, which couple only very weakly to the particles of the Standard Model (SM). The simplest possible interactions between these two sectors have been systematically classified within the so-called \textit{portal framework}~\cite{Wilczek:1977pj,Weinberg:1977ma,Shifman:1979if,Zhitnitsky:1980tq,Dine:1981rt,Holdom:1985ag,Dodelson:1993je,Boehm:2003hm,Asaka:2005an,Asaka:2005pn,OConnell:2006rsp,Patt:2006fw,Pospelov:2007mp,Gorbunov:2007ak,Batell:2009di,Krnjaic:2015mbs,Arina:2021nqi,Manzari:2022iyn}, giving rise to several benchmark scenarios for SM interactions with a dark sector that can be tested  experimentally (see Refs.~\cite{Alekhin:2015byh,Battaglieri:2017aum,Beacham:2019nyx,Lanfranchi:2020crw,Krnjaic:2022ozp} for reviews). 

Interestingly, if the dark  particles are sufficiently light to be produced in stellar plasmas, then their emission modifies the standard picture of stellar evolution and stringent constraints on SM interactions with the dark sector can be obtained also from astrophysical observations~\cite{Sato:1975vy,Dicus:1978fp,Vysotsky:1978dc,Dicus:1979ch,Raffelt:1996wa,Balaji:2022noj,Lella:2022uwi,Turner:1991ax,Raffelt:1993ix,Caputo:2021rux,Caputo:2022rca}. 
Ordinary stars yield strong constraints on DM coupled to electrons and photons~\cite{Raffelt:1996wa}, while the extreme temperatures and densities reached in the proto-neutron stars (PNS) formed during core-collapse supernovae (SN) allow one to probe also DM couplings to nucleons~\cite{Raffelt:1987yt,Turner:1987by,Mayle:1987as,Burrows:1988ah,Burrows:1990pk,Dent:2012mx,Rrapaj:2015wgs,Chang:2018rso,Carenza:2019pxu,Sung:2021swd,Calore:2021klc,Choi:2021ign,Shin:2021bvz,Calore:2021lih,Lucente:2022vuo,Balaji:2022noj,Lella:2022uwi}, pions~\cite{Turner:1991ax,Raffelt:1993ix,Keil:1996ju,Carenza:2020cis,Fischer:2021jfm,Shin:2022ulh}, hyperons~\cite{MartinCamalich:2020dfe,Camalich:2020wac,Vonk:2021sit} and muons~\cite{Bollig:2020xdr, Calibbi:2020jvd, Croon:2020lrf, Caputo:2021rux}. Most of these analyses focus on direct production and emission of light dark bosons (such as axions or dark photons) from the stellar medium.

In this article we aim to study instead the case where these bosons merely serve as   massive mediators between dark fermions and SM leptons; i.e. they serve only as a portal to the dark sector, which allows the production of sufficiently light dark fermions in stellar plasmas. Prominent examples of this scenario are gauged lepton flavor models such as $U(1)_{L_{\mu} - L_{\tau}}$, which contain a light massive gauge boson~\cite{He:1990pn,He:1991qd,Ma:2001md,Baek:2001kca,Salvioni:2009jp,Heeck:2011wj,Feng:2012jn,Harigaya:2013twa,Bauer:2018onh,Amaral:2021rzw,Greljo:2021npi,Greljo:2022dwn,Langacker:2008yv} and a dark sector charged under the corresponding group~\cite{Foldenauer:2018zrz,Asai:2020qlp,Holst:2021lzm,Drees:2021rsg,Hapitas:2021ilr,Carpio:2021jhu,Heeck:2022znj,Qi:2022kgs,Deka:2022ltk,Baek:2022ozm,Escudero:2022gez,Hooper:2023fqn}. This type of scenarios has attracted much attention as they can simultaneously address the $(g-2)_\mu$~ anomaly~\cite{Aoyama:2020ynm,Muong-2:2006rrc,Muong-2:2021ojo,Borsanyi:2020mff,Alexandrou:2022amy,FermilabLattice:2022izv,Ce:2022kxy}, provide a DM candidate with the right abundance, and contribute to the effective number of neutrino species, alleviating the Hubble constant tension~\cite{Planck:2018vyg,Brout:2022vxf,Bernal:2016gxb,Escudero:2019gzq,DiValentino:2021izs}. Although we will present novel constraints on this scenario from SN~1987A later on, we perform our analysis within a more general setup.

For definiteness, let us consider a vector mediator $Z^\prime$ with mass $m_{Z^\prime}$ and couplings to SM leptons and dark fermions $\chi$ with mass $m_\chi$, 
\begin{align}
{\cal L} & \supset   Z_\nu^\prime \left( g_\ell \overline{\ell} \gamma^\nu \ell + g_{\nu_\ell} \overline{\nu_\ell} \gamma^\nu P_L \nu_{\ell} + g_\chi \overline{\chi} \gamma^\nu \chi \right) \, ,
\label{eq:LagSimplified}
\end{align}
where $\ell = e, \mu,\tau$, and $g_\ell$, $g_{\nu_\ell}$ and $g_\chi$ are generic couplings. 
Assuming that the dark fermions are sufficiently light ($m_\chi \lesssim 150 \, \MeV$), 
we will show in the following that their production from the PNS in SN~1987A leads to stringent constraints on their couplings to SM leptons. For the benchmark $U(1)_{L_{\mu} - L_{\tau}}$ model, this excludes large regions of the parameter space targeted by future experiments~\cite{Alekhin:2015byh,Krnjaic:2019rsv,Kahn:2018cqs,Galon:2019owl,Asai:2021wzx,Sieber:2021fue,Bandyopadhyay:2022klg,Cesarotti:2022ttv,Moroi:2022qwz,NA64:2022rme,Rella:2022len,Ariga:2023fjg}.

While our analysis is valid for any mass of the $Z^\prime$, we can integrate it out and describe its contribution with an effective  
four-fermion operator, if $\mZ$ is much larger than the PNS's temperature and chemical potentials. This leads to significant simplifications in the analysis, and allows us to extend it to derive SN~1987A bounds on completely generic interactions of the dark sector with leptons through heavy portal mediators. In the heavy $Z^\prime$ limit, our calculations are in fact analogous to the ones necessary to study the SM production of neutrinos from leptons in stellar plasmas, which have received much attention over the past half century, since the seminal works in the 1960's~\cite{Chiu:1961zza,ritus_1961,1963PhRv..129.1383A,Petrosian:1967alk,1967ApJ...150..979B,Dicus:1972yr,1976PhRvD..13.2700S,1985ApJ...296..197M,1987ApJ...313..531S,1989ApJ...339..354I,1992ApJ...392...70B,1993PhRvD..48.1478B,1994ApJ...425..222H,1996ApJS..102..411I,2003NuPhB.658..217E,2004PhRvD..69b3005D,2007MNRAS.381.1702K} (see also Refs.~\cite{1976PhRvD..13.2700S,1977PhRvD..15..977D,Barbieri:1988av,Guha:2018mli} for production of light dark fermions from heavy new physics). On the other hand, in case the $Z^\prime$ is a light and narrow state, the calculation is similar to the on-shell production of massive vector bosons coupled to leptons in the plasma~\cite{Croon:2020lrf,Kachelriess:2000qc,Lai:2022csw,Akita:2022etk,2023arXiv230412907L} (see also Refs.~\cite{Bollig:2020xdr, Croon:2020lrf, Caputo:2021rux,Lucente:2021hbp} for massive axions). 

In this article we focus on the portal interactions with muons, but we also study neutrinos, which could be naturally linked to muons via $SU(2)_L$. Electrons however form a highly degenerate and ultra-relativistic plasma in the PNS, which might lead to important medium effects in the electron and photon dispersion relations, requiring the inclusion of other production mechanisms not relevant for muons and neutrinos. With this caveat in mind, our calculations are easily extensible to electrons, generalizing and updating the pioneering work of Ref.~\cite{Barbieri:1988av} and improving the results presented in Ref.~\cite{Guha:2018mli}. 

The rest of the paper is organized as follows. In Sec.~\ref{sec:SN} we outline the classical SN argument to constrain new exotic cooling agents using the neutrino flux observed from SN~1987A. Besides describing the general theoretical framework, we specify the SN simulations that we employ in our numerical analysis. In Sec.~\ref{sec:production} we describe and compute the rates of the main emission mechanisms induced by the model in Eq.~\eqref{eq:LagSimplified}. We focus on extracting the main physical features of the rates using  different approximations in the various regimes of the $Z^\prime$ mass, and on deriving analytical estimates. However, our final results rely on exact numerical computations whose details are deferred to Appendices.  In Sec.~\ref{sec:Results}, then, we implement these  calculations of the rates in the SN simulations and derive the constraints on the parameter space of the $Z^\prime$ model in Eq.~\eqref{eq:LagSimplified}. We also generalize this analysis in terms of effective operators in the heavy $Z^\prime$ limit, and to one particular realization of the model arising from a $L_\mu-L_\tau$ gauged symmetry. Finally, in Sec.~\ref{sec:Conclusions} we summarize the results of our paper and close with a brief outlook.

\section{Supernova Cooling}
\label{sec:SN}

In the dense and hot environment within proto-neutron stars~\cite{Burrows:1986me,Bethe:1990mw,1997PhDT........18R,Janka:2012wk} neutrinos become trapped and a thermal population of muons is predicted to arise~\cite{Bollig:2017lki,Fischer:2020vie}. 
New light dark particles that couple to  leptons, e.g. via the interactions in Eq.~\eqref{eq:LagSimplified}, can be produced efficiently in the stellar plasma, leading to a significant loss of energy if they can escape from the PNS. The corresponding dark luminosity $L_\chi$ is then subject to the classical bound $L_\chi\lesssim L_\nu$ at $1\,$s post-bounce, where $L_\nu$ is the neutrino luminosity~\cite{Raffelt:1996wa,Burrows:1988ah,Burrows:1990pk}. This limit is obtained from the observation of a neutrino pulse over $\sim 10\,$s~\cite{Loredo:2001rx,Vissani:2014doa} during SN~1987A~\cite{Bionta:1987qt,Hirata:1987hu,Alekseev:1988gp}, which is in accordance with the predictions of the standard theory of core-collapse SN  (see Refs.~\cite{Bar:2019ifz,Page:2020gsx,Bollig:2020xdr} for a critical reappraisal of  this limit\footnote{It has also been noted in Ref.~\cite{Li:2023ulf} that there is a disagreement between the results of simulations and the observed neutrino signal of SN~1987A during the first second. However, this conclusion has been disputed in Ref.~\cite{Fiorillo:2023frv}, where a more extensive analysis with a wider time window has been performed.}).  

Here, we apply this argument to scenarios where light dark fermions couple to leptons with interactions such as those in Eq.~\eqref{eq:LagSimplified}. One needs to distinguish two regimes based on the mean free path (MFP) of the dark fermions in the plasma or, equivalently, the strength of the portal interactions. If the  dark particles are very weakly coupled (or the MFP is much larger than the radius of the PNS) then they \textit{free stream} out from the SN once produced, whereas for large couplings (or MFP much shorter than the radius of the PNS) they thermalize with the medium and get \textit{trapped} inside of the PNS. 

In the free-streaming regime the general expression for the total energy-loss rate per unit volume, $Q$, for a given emission process is
\begin{align}
Q=& \int \Big[\prod_{\text{init},\,i} \frac{d^3p_i}{(2\pi)^3 2E_i} f_i \Big] \Big[ \prod_{\text{final},\,j} \frac{d^3p_j}{(2\pi)^3 2E_j} (1\pm f_j)\Big] \nonumber\\
&\times (2\pi)^4 \delta^4 (\sum_i p_i-\sum_j p_j) \sum_{\rm spins}|\mathcal{M}|^2 E_\chi \, . 
\label{eq:FSgeneral}
\end{align}
These are thermal integrals over the phase space of all the initial- and final-state particles  weighted by their number density distributions $f_i$ and the Pauli blocking or Bose enhancement factors $(1\mp f_j)$, respectively. Furthermore, $|\mathcal{M}|^2$ is the the squared matrix element of the given production process and $E_\chi$ is the total energy carried away by the dark particles. In the calculations of the free-streaming regime, one conventionally uses $f_\chi =0$ for the new particles, because their occupation numbers inside the PNS are very low and not thermalized by assumption.

In the deep trapping regime, on the other hand, the dark-sector particles are in thermal equilibrium with the plasma and they are emitted from a surface with radius $r_\chi$ (\textit{dark sphere}) following a law analogous to the one of the black body radiation, 
\begin{align}
L_{\chi}^{\rm trap} = \frac{\mathfrak g_\chi}{\pi} r_\chi^2 T_\chi^4 \int_{x_m}^{\infty} dx \frac{x^2 \sqrt{x^2 - x_m^2}}{e^x + 1} \,,
\label{eq:blackbody}
\end{align}
where $\mathfrak g_\chi$ is the number of degrees of freedom of the $\chi$ particle ($\mathfrak g_\chi=2$ for massive dark fermions), $x_m = m_\chi /T_\chi$ and $T_\chi = T( r_\chi)$ is the temperature of the dark sphere. The radius $r_\chi$ is defined, as is conventional in astrophysics~\cite{1967pswh.book.....Z,1967aits.book.....C,2019leas.book.....W}, through the optical depth $\tau_\chi (r)$, by requiring 
\begin{align}
\tau_\chi (r_\chi) = \int_{r_\chi}^\infty \frac{dr}{\lambda(r)}= \frac{2}{3} \, ,
\label{eq:OpDepth}
\end{align} 
where $\lambda(r)$ is a suitable spectral average of the dark fermion's MFP at a radius $r$. In this work, we use a ``naive'' thermal average
\begin{align}
\lambda (r)  = \langle \lambda (r, p_\chi)\rangle_\chi \equiv \frac{\mathfrak g_\chi}{n_\chi (r)} \int \frac{d^3 p_{\chi}}{(2 \pi)^3} \frac{\lambda (r, p_\chi)}{e^{E_\chi/T(r)} + 1} \,,
\label{eq:thermalaverage}
\end{align}
for computational simplicity\footnote{We have checked that other averages, such as the conventional Rosseland MFP~\cite{1967pswh.book.....Z,1967aits.book.....C,2019leas.book.....W}, give very similar results. In the recent literature also other forms of taking the spectral average have been discussed, for example by including an energy-dependent opacity in the free-streaming spectral luminosity, integrating over the energy only in the very end~\cite{Croon:2020lrf}. For a discussion and comparison of the various approaches see Ref.~\cite{Caputo:2021rux}.}  (see Appendix~\ref{app:absorption} for our definitions of thermal averages). The energy-dependent MFP $\lambda (r, p_\chi)$ is related to the \textit{total} rate of interaction of a dark-sector particle in the medium, $\Gamma_\chi=v_\chi/\lambda (r, p_\chi)$, through its velocity $v_\chi=p_\chi/E_\chi$. 

The contribution to $\Gamma_\chi$ of a given process with a bunch of target particles $b$ colliding with  $\chi$ in the initial state is defined through
\begin{equation}
\mathcal{C}_{\rm abs}^b = \mathfrak g_\chi \int \frac{d^3p_\chi}{(2\pi)^3} f_\chi \Gamma_\chi^b \, .
\label{eq:gamma_chi}
\end{equation}
The quantity $\mathcal C_{\rm abs}$ is the collision operator describing the absorption rate per unit volume of the medium
\begin{align}
\mathcal C_{\rm abs}^b=& \int \Big[\prod_{\text{init},\,i}\frac{d^3p_i}{(2\pi)^3 2E_i} f_i \Big] \Big[ \prod_{\text{final},\,j} \frac{d^3p_j}{(2\pi)^3 2E_j} (1\pm f_j)\Big] \nonumber\\
&\times (2\pi)^4 \delta^4 (\sum_i p_i-\sum_j p_j) \sum_{\rm spins}|\mathcal{M}|^2] \, , 
\label{eq:Coll_gen}
\end{align}
which uses the same definitions as in Eq.~\eqref{eq:FSgeneral}, except for $|\mathcal{M}|^2$ which is now the squared matrix element of the given \textit{absorption} process\footnote{Denoting $\Gamma_\text{ours}$ for our definition, the standard approach in the literature~\cite{Caputo:2021rux} is to work in terms of \emph{emission} rates $\Gamma_E = \mathfrak{g}_\chi f_\chi \Gamma_\text{ours}$. However, one then has to use Boltzmann-equation arguments to argue which rate has to be used for absorptive processes, leading to the \emph{reduced absorption} rate $\Gamma_A$. The definition $\Gamma_\text{ours}$ is chosen such that $\Gamma_\text{ours}=\Gamma_\chi^{b} = n_b \langle \sigma v\rangle_b$, which naturally leads to $\Gamma_\text{ours} = \Gamma_A$ (see Appendix~\ref{app:absorption}).}.

For the numerical analyses of this paper we use SN simulations including muons presented in Ref.~\cite{Bollig:2020xdr} and whose radial profiles for the relevant quantities are reported in~\cite{Garching}. Our fiducial results are obtained using the simulation labeled as SFHo-18.80, which reaches the lowest temperatures and, therefore, will lead to the most conservative limits on the dark luminosity (at 1\,s post-bounce). The upper bound is set by the neutrino luminosity calculated within the  same simulation, which for SFHo-18.80 is given by\footnote{We use the total co-moving neutrino luminosities reported in the simulations at the radius of the neutrino-sphere $\sim16$ km. Note that including relativistic corrections would reduce both the neutrino and the dark fermion flux as seen by a distant observer by roughly 30\%~\cite{Caputo:2021rux}. We thank R. Bollig and H-.~T.~Janka for facilitating us the necessary data to make these estimates. See also Ref.~\cite{Caputo:2021rux}.}
\begin{align}
L_\chi\le L_\nu=5.7\times10^{52}\text{ erg s}^{-1} \, .
\label{eq:Lchi}
\end{align}
For a rough estimate of the systematic uncertainties related to SN modeling, we will also show the more stringent limits obtained from using the hotter SFHo-20.0 simulation, which gives $L_\chi\le
1.0\times10^{53}\text{ erg s}^{-1}$. In the free-streaming regime, the dark luminosity is obtained as a volume integral of Eq.~\eqref{eq:FSgeneral}, $L_\chi=\int Q dV$, while in the trapping regime we use Eq.~\eqref{eq:blackbody}. 

Finally, it will be useful to estimate the contributions to the dark luminosity of the different processes to understand their relative importance. For this, we define ``typical PNS conditions'' as those at 1\,s post-bounce and at a radius $\approx 10$ km. This region dominates the volume emission in $L_\chi$ and is representative of the bounds in the free-streaming regime. Using the simulation SFHo-18.80~\cite{Garching}, this approximately corresponds to:
\begin{equation}
\boxed{
\arraycolsep=1.4pt\def\arraystretch{1.5}
\begin{array}{c}
\text{\bf Typical PNS conditions}\\
\;\;\;T = 30 \, \MeV,\;\rho=2\times10^{14}$ g cm$^{-3},\;\;\;\\
\;\;\;\mu_\mu =100 \, \MeV,\; \mu_{\nu_e} =20 \, \MeV,\;\;\;\\
\;\;\;\mu_{\nu_\mu} =-10 \, \MeV,\; \mu_e = 130 \, \MeV,\;\;\;\\
Y_\mu=0.026, \;\;\;Y_e=0.12. 
\end{array}
}
\label{eq:typicalPNS}
\end{equation}
Here $T$ denotes the temperature, $\rho$ the density, $\mu_l$ the chemical potential of the lepton $l$ and $Y_\ell$ is the number density fraction of the charged lepton $\ell$ relative to the one of baryons. For the $Y_\ell$ we quote the results derived from the rounded temperature and chemical potentials in Eq.~\eqref{eq:typicalPNS} and, therefore, they are slightly different to those reported in~\cite{Garching}. Let us stress again that Eq.~\eqref{eq:typicalPNS} will be only used for numerical estimates, while our final results and constraints on the models will be obtained using the full radial profiles of all relevant thermodynamical quantities. 

\section{Production and absorption rates}
\label{sec:production}

\begin{figure}[t]
	\centering
	\includegraphics[width=0.5\columnwidth]{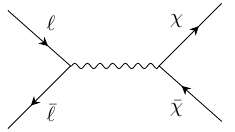}\\
    \includegraphics[width=0.4\columnwidth]{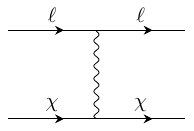}\\
    \includegraphics[width=0.7\columnwidth]{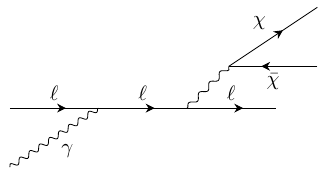}
	\caption{Most relevant $\ell-\chi$ processes for supernova cooling where $\ell$ is a charged lepton. Top-to-bottom panels represent annihilation, scattering and photoproduction processes, respectively. Annihilation and scattering diagrams will also contribute when $\ell$ is replaced by a neutrino. }
	\label{fig:FeynmanDiagrams}
\end{figure}

There are two main production mechanisms of $\chi\bar\chi$ pairs from muons and neutrinos in SN (see top and bottom panel of Fig.~\ref{fig:FeynmanDiagrams}): \textit{(i)} Annihilation $\mu^-\mu^+\to \chi\bar\chi$ and $\nu_\ell\bar\nu_\ell\to\chi\bar\chi$; \textit{(ii)} photoproduction $\gamma\mu^-\to\mu^-\chi\bar\chi$. We do not consider bremsstrahlung processes, $\mu^-p\to\mu^- p\chi\bar\chi$, because they are suppressed with respect to photo-production (or semi-Compton production) of (pseudo)scalars and $Z^\prime$~\cite{Croon:2020lrf,Bollig:2020xdr,Caputo:2021rux}. 

Note also that this is not generally true for production from electrons because they are ultra-relativistic and form a highly degenerate system that suppresses photoproduction compared to bremsstrahlung and annihilation~\cite{Raffelt:1996wa}. Moreover, there are important plasma effects which, for example, dress the electron with an effective mass $m_e^*\sim 10\text{ MeV}$ and give rise to pseudo-particle excitations that need to be taken into account in a realistic analysis (see e.g. Ref.~\cite{Lucente:2021hbp} for the emission of massive axions from electrons in SN). Nonetheless, the calculations we present in this work can be easily extended to electrons and compared with previous literature where all these effects have been neglected~\cite{Barbieri:1988av,Guha:2018mli}. 
We will estimate some of them below in Sec.~\ref{sec:uncertainties}. 

In case of absorption, there are the inverse processes $\mu^-\chi\bar\chi\to \gamma\mu^-$ and $\chi\bar\chi\to\mu^-\mu^+$, whose rates are related by detailed balance to those of the photoproduction and annihilation production, respectively, provided that the $\chi$ and $\bar \chi$ particles reach thermal equilibrium. In addition, other scattering processes may contribute to the diffusion and energy transport in the trapping regime, such as $\chi\mu^-\to \chi\mu^-$ and $\chi\nu_\ell\to\chi\nu_\ell$ (see middle panel of Fig.~\ref{fig:FeynmanDiagrams}), or processes in the dark sector such as $\chi\bar\chi\to\chi\bar\chi$. 

In Appendices~\ref{app:2to2} and \ref{app:2to3} we provide the cross sections for all relevant $2 \to 2$ and $2 \to 3$ processes needed to calculate the energy-loss and absorption rates. In the following, we discuss in detail the contributions of the annihilation and photoproduction topologies. 

\subsection{Annihilation}
\label{sec:annihilation}

The energy-loss rate per unit volume in Eq.~\eqref{eq:FSgeneral} for $\mu^-\mu^+\to\chi\bar\chi$ annihilation can be simplified to
\begin{align}
\label{eq:Qann}
Q & =\frac{\mathfrak g_\mu^2}{16 \pi^4}\int_{m_\mu}^\infty d E_+\int_{m_\mu}^\infty   d E_- \left( E_+ + E_- 
\right)  \overline{p}_+ \overline{p}_- f_+  f_- I_s  \,.
\end{align}
In this equation, $\mathfrak g_\mu$ are the  muon's spin degrees of freedom, $E_{-(+)}$ denote the muon (anti-muon) energy in the PNS rest frame, $\overline{p}_{\pm}$ are the absolute values of their 3-momenta, $\overline{p}_\pm \equiv \sqrt{E_{\pm}^2 - m_\mu^2}$, and  $f_{{\pm}} \equiv f_{{\pm}} (E_\pm)= 1/(e^{(E_{\pm} \pm \mu)/T}+1)$ are their Fermi-Dirac distributions in the medium. The function $I_s$ is a (dimensionless) angular integral over the annihilation cross section $\sigma (s) = \sigma ({\mu^+ \mu^- \to \chi \overline{\chi}})$
\begin{align}
\label{eq:I_function}
 I_s = \int d \cos \theta \, s \sqrt{1 -4m_\mu^2/s} \,  \sigma(s)   \, ,
\end{align}
which depends on the colliding angle $\theta$ between the muon and anti-muon through the Mandelstam variable $s$ in the PNS frame, $s = 2 \left( m_\mu^2 + E_+ E_- -\overline{p}_+ \overline{p}_-  \cos \theta \right)$. The cross section is physical only above the 2-particle threshold which imposes the kinematic constraint $s\geq 4 \max ({m_\chi^2,  m_\mu^2})$ in the angular integral. 

The annihilation cross section in Eq.~\eqref{eq:I_function} for the $Z^\prime$ model in Eq.~\eqref{eq:LagSimplified} is
 \begin{align}
\sigma(s) = \frac{g_\mu^2 g_\chi^2 }{3\mathfrak g_\mu^2 \pi} \frac{s}{(s-m_{Z^\prime}^2)^2 + m_{Z^\prime}^2 \Gamma_{Z^\prime}^2  }
\frac{\beta_\chi(s)}{\beta_\mu(s)} \kappa_\mu(s) \kappa_\chi(s) \, ,
\label{eq:sigma_annZp}
\end{align}
where we have introduced $\beta_i(s)=\sqrt{1-4m_i^2/s}$, $\kappa_i(s)= 1 + 2 m_i^2/s$ and the total $Z^\prime$ width 
\begin{align}
\Gamma_{Z^\prime} = \frac{m_{Z^\prime}}{12 \pi }  \sum_{i} g_i^2 \kappa_i(m_{Z^\prime}^2)\beta_i(m_{Z^\prime}^2)\theta(m_{Z^\prime}-2m_i)  \,,
\label{eq:Zp_width}
\end{align}
and where $\theta(x)$ is the Heavyside step function. 

Note that the average energy carried away by the dark fermions in the annihilation process is equal to the thermally averaged center-of-mass (CM) energy of the leptons,
\begin{align}
\label{eq:AvEnergyAnn}
\mathcal E_l \equiv \sqrt{\langle s \rangle_{l \bar{l}}}.
\end{align}
For typical conditions in the PNS, Eq.~\eqref{eq:typicalPNS}, $\mathcal E_\mu\sim 280 \, \MeV$, which sets the scale for $2 m_\chi$ above which the production of $\chi$'s in the plasma will become exponentially (``Boltzmann") suppressed by the distribution functions $f_\pm$. 
In addition, this energy scale allows one to define three regimes of $m_{Z^\prime}$ in Eq.~\eqref{eq:sigma_annZp} depending on which term dominates the denominator: \textit{(i)} A ``{\bf heavy regime}" in which $m_{Z^\prime}  \gtrsim 1\,\GeV \gg  \mathcal E_\mu$, so that the $Z^\prime$ is too heavy to be produced on-shell; \textit{(ii)} the ``{\bf resonant regime}" where the $Z^\prime$ can be produced on-shell, $2 m_\mu \le m_{Z^\prime}  \lesssim  \mathcal E_\mu$; and \textit{(iii)} the ``{\bf light regime}'' with $Z^\prime$ masses below the two-muon threshold, $m_{Z^\prime}  < 2 m_\mu$, so that the $Z^\prime$ is too light to be produced on-shell. 

Analogous expressions can be defined for electrons and neutrinos by replacing the couplings and masses accordingly (notice that for neutrinos $\mathfrak g_{\nu_\ell}=1$). Moreover, the average CM energies in these cases are $\mathcal E_e \sim 160 \, \MeV$ and $\mathcal E_\nu \sim 130 \, \MeV$, and analogous regimes to those for the muons can be formulated for neutrino-antineutrino and electron-positron annihilation. The light regime in these cases is restricted to extremely small $Z^\prime$ masses, making it irrelevant for the range of vector boson masses we are considering here. 
 
The demarcation of these regimes is useful because one can use approximations to derive analytic results and isolate the main physical factors in control. In the following, we discuss these approximations and describe their contributions to the absorption rate $\Gamma_\chi$.

\subsubsection{Heavy regime}
\label{sec:AnnHeavy}

In this case, the denominator in the propagator (see Eq.~\eqref{eq:sigma_annZp}) is dominated by the $Z^\prime$ mass. Expanding in powers of $s/m_{Z^\prime}^2$ up to leading order, the cross section can be easily integrated analytically, giving a function $I_s (E_+, E_-, m_\chi, m_\mu)$ proportional to the effective coupling $g_\chi^2 g_\mu^2/\mZ^4$. We can further approximate this expression by taking the high-energy limit $m_\chi\to0$ and $m_\mu \to 0$, obtaining\footnote{In the following discussion we fix $\mathfrak g_\mu=2$ with the understanding that some intermediate formulas change by factors of 2 for the neutrino case.}
\begin{align}
I_s^{\rm heavy} (E_+, E_-, 0, 0) & = \frac{8 g_\chi^2 g_\mu^2}{9 \pi } \frac{ E_+^2 E_-^2 }{\mZ^4} \, .
\label{eq:Is_heavy}
\end{align}
Also setting $m_\mu \to 0$ in the integrals in Eq.~\eqref{eq:Qann}, the integrations can be carried out analytically, giving
 \begin{align}
 Q^{\rm heavy} & = \frac{2 g_\chi^2 g_\mu^2}{9 \pi^5 } \frac{ T^9 }{\mZ^4} \left[
H_4(y)H_3(-y)+(y \to -y) \right] \, . 
\label{eq:Q_ann_heavy}
 \end{align}
Here we have re-scaled the chemical potential $y=\mu/T$ and introduced the functions 
\begin{align}
\label{eq:H_function} 
H_n(y)=\int_{0}^{\infty} dx\frac{x^n}{e^{x-y}+1}=- n! \, {\rm Li}_{n+1}(-e^{y})\,,
\end{align}
where ${\rm Li}_{n+1}(z)$ is the polylogarithm of order $n+1$. If we also take vanishing chemical potentials, we recover the results in Ref.~\cite{Barbieri:1988av}
\begin{align}
\label{eq:Q_ann_heavy_massless}
Q^{\rm heavy}_0 = \frac{4 g_\chi^2 g_\mu^2}{9 \pi^5 } \frac{T^9}{m_{Z^\prime}^4}F_4\,F_3   \, , 
\end{align}
in terms of the Riemann $\zeta$-function
\begin{align}
\label{eq:Fn}
F_n=H_n(0)=n! (1-2^{-n})  \zeta(n+1) \, ,
\end{align}
with $F_4 F_3 \approx 133$. We have also included a subindex in $Q$ to indicate that this is a zeroth-order approximation neglecting masses and chemical potentials of the leptons. 

In order to assess the accuracy of the above approximations, we compare $Q$ in Eq.~\eqref{eq:Qann} for massless dark fermions and the cross section in the heavy $Z^\prime$ limit with Eq.~\eqref{eq:Q_ann_heavy_massless} for different SM particles at the typical conditions of PNS in Eq.~\eqref{eq:typicalPNS}. For muons one finds $Q_\mu/Q_0 \approx 0.33$ while for neutrinos and electrons one finds $Q_\nu/Q_0 \approx 0.99$ and $Q_e/Q_0 \approx 0.54$ (using the physical electron mass in vacuum), respectively. 

The thermal suppression of the muon population is mild for these conditions in the PNS, $Y_e\simeq4 Y_\mu$. The positron abundance is also suppressed by the large electronic chemical potential and, hence, for the same couplings to electrons and muons one obtains similar rates. 

With these approximations one can estimate the parametric dependence of the energy loss rate per unit mass (i.e.~the emissivity)  produced by lepton annihilation in the heavy regime as
\begin{align}
\label{eq:emiss_ann_heavy}
\epsilon_{\rm ann}^{\rm heavy} =  \epsilon_{\rm max}  \left( \frac{T}{30 \, \MeV}  \right)^9 \left(\sqrt{g_\chi g_l}\;\frac{4.1 \, \TeV}{\mZ} \right)^4 \, , 
\end{align}
where we have divided Eq.~\eqref{eq:Q_ann_heavy_massless} by the density in Eq.~\eqref{eq:typicalPNS}, and $\epsilon_{\rm max}=2.1\times10^{19}$ erg s$^{-1}$ g$^{-1}$ has been estimated dividing $L_\chi$ in~\eqref{eq:Lchi} by the total mass of the PNS in this simulation $M_{\rm PNS}=1.351~M_{\odot}$.

\subsubsection{Resonant regime}
\label{sec:AnnResonant}

If the $Z^\prime$ can be produced on-shell, then the denominator in Eq.~\eqref{eq:sigma_annZp} is dominated by the $Z'$ width $\Gamma_{Z^\prime}$, and it can be replaced by $\pi/(\mZ \Gamma_{Z^\prime}) \delta (s-\mZ^2)$ in the narrow width approximation. The $\delta$-function can be used to perform the angular integration in Eq.~\eqref{eq:I_function} and for $m_\chi \to 0$  this gives (neglecting terms of relative size $2 m_\mu^2/\mZ^2$)
\begin{align}
I_s^{\rm res} & = \frac{ g_\chi^2 g_\mu^2}{24  } \frac{\mZ^3}{\Gamma_{Z^\prime} E_+ E_-} \,.
\label{eq:Is_res}
\end{align}
The energy integrations of Eq.~\eqref{eq:Qann} can be well-approximated by neglecting the chemical potentials, but keeping a non-zero muon mass in the integrand, giving
\begin{align}
 Q^{\rm res}_{\mu =0} & = \frac{ g_\mu^2 {\rm BR}_{\chi} }{4 \pi^3 } \mZ^2  T^2  
m_\mu e^{-2 m_\mu/T} \, ,
\label{eq:Q_ann_res_mu0}
 \end{align}
where $ {\rm BR}_{\chi} \equiv {\rm BR} (Z^\prime \to \chi \overline{\chi})$ denotes the invisible $Z^\prime$ branching ratio, and numerically  $ m_\mu/T  e^{-2 m_\mu/T} \approx 0.004$. For muon annihilation this indeed yields a good approximation, with $Q_\mu/Q_{\mu=0} \approx 0.95$, while for electrons one can set $m_e \to 0$ in the integrals, giving a result similar to Eq.~\eqref{eq:Q_ann_heavy}, which can be further approximated by\footnote{For $y \gg 1$, one has $H_n(y) \approx y^{n+1}/(n+1), H_n(-y) \approx e^{-y} n! $. } 
\begin{align}
 Q^{\rm res}_{m =0} & = \frac{ g_\mu^2 {\rm BR}_{\chi} }{16 \pi^3 } \mZ^2  T \mu_e^2 e^{-\mu_e/T} 
 \, ,
\label{eq:Q_ann_res_m0}
 \end{align}
 and numerically  $ \mu_e^2/4T^2  e^{- \mu_e/T} \approx 0.06$, resulting in $Q_e/Q_{m=0} \approx 1.6$. Finally 
neglecting lepton masses and chemical potential  simplifies to  \begin{align}
\label{eq:Q_ann_res_massless}
Q^{\rm res}_0 = \frac{ g_\mu^2{\rm BR}_\chi}{4 \pi^3 }m_{Z^\prime}^2T^3\,   F_1 F_0   \, ,
\end{align}
where numerically $F_1 F_0 \approx 0.57$. This is a good approximation for neutrino annihilation with $Q_\nu/Q_0 \approx 0.95$, while energy-loss rates for electrons (muons) are smaller by a factor 10 (100).   

Importantly, the contribution to the energy-loss rates in the resonant regime scales perturbatively with the couplings as $\sim\mathcal O(g^2)$ instead of $\sim\mathcal O(g^4)$ in the heavy or light regimes. In fact, for ${\rm BR}_{\chi}=1$, one should recover the results obtained for the coalescence $Z^\prime$ production mechanisms in Ref.~\cite{Croon:2020lrf}\footnote{Indeed, using the approximation $f_+ (E_+) f_- (E_-) \approx f_{Z^\prime} (E_+ + E_-)$, we reproduce their expression for the $Z^\prime$ production rate  in the light $Z^\prime$ and for massless leptons and $\chi$, up to a factor $2/3$. 
See also footnote {\color{red}3} in Ref.~\cite{Caputo:2021rux} for a discussion of the possible origin of these discrepancies.}. Also, the various rates in Eqs.~\eqref{eq:Q_ann_res_mu0}, \eqref{eq:Q_ann_res_m0} and \eqref{eq:Q_ann_res_massless} all scale quadratically with the $Z^\prime$ mass.

From these expressions one can readily obtain the emissivities, which for  neutrino annihilation read 
\begin{align}
\label{eq:emis:ann:res}
\epsilon_{\rm ann,\nu}^{\rm res} =   \epsilon_{\rm max} \left( \frac{T}{30 \, \MeV}  \right)^3\left(\frac{g_{\nu_\ell} }{10^{-9}}\;\frac{m_{Z^\prime}}{10\;\MeV}\right)^2  {\rm BR}_{\chi}   \, ,
\end{align}
where we have used the same approximations as in Sec~\ref{sec:AnnHeavy} which are valid up to $m_{Z^\prime}\sim200$ MeV. As discussed above, emissivities for electrons and muons are expected to be smaller by a factor 10 and 100, respectively.

\subsubsection{Light regime}   

In this case the denominator of the propagator is dominated by $s$, and the cross section can be integrated analytically. For massless $\chi$ and muons one obtains 
\begin{align}
I_s^{\rm light} & = \frac{ g_\chi^2 g_\mu^2}{6 \pi }  \, ,
\label{eq:Is_light}
\end{align}
which is independent of $E_+, E_-$. Similarly as in the resonant regime, the energy integrations of Eq.~\eqref{eq:Qann} can be approximated by neglecting the chemical potentials, but keeping a non-zero lepton mass in the integrand
\begin{align}
 Q^{\rm light}_{\mu =0} & = \frac{ g_\chi^2 g_\mu^2}{12 \pi^5 }  T^2  
m_\mu^3 e^{-2 m_\mu/T} \, ,
\label{eq:Q_ann_lighy_mu0}
 \end{align}
where  $m_\mu^3/T^3  e^{-2 m_\mu/T} \approx 0.04$. For muon annihilation this indeed yields a good approximation, with $Q_\mu/Q_{\mu=0} \approx 0.85$, while for electrons one can set $m_e \to 0$ in the integrals (but keep non-zero chemical potentials), giving a result similar to Eq.~\eqref{eq:Q_ann_res_m0}
\begin{align}
 Q^{\rm light}_{m_\ell =0} & = \frac{ g_\chi^2 g_\mu^2  }{72 \pi^5 }  T^2 \mu_e^3 e^{-\mu_e/T} 
 \, ,
\label{eq:Q_ann_light_m0}
 \end{align}
 and numerically  $ \mu_e^3/6T^3  e^{- \mu_e/T} \approx  0.2$, resulting in $Q_e/Q_{m=0} \approx 1.2$. Finally 
neglecting lepton masses and chemical potential  simplifies to  
\begin{align}
\label{eq:Q_ann_light_massless}
Q^{\rm light}_0 = \frac{ g_\chi^2 g_\mu^2}{12 \pi^3 }  T^5 F_2 F_1   \, ,
\end{align}
where numerically $F_2 F_1 \approx 1.5$. This is a good approximation for neutrino annihilation with $Q_\nu/Q_0 \approx 0.97$, while for electrons (muons) the energy-loss rates and emissivities are smaller by a factor 10 (100) than predicted by this formula. Nevertheless, using Eq.~\eqref{eq:Q_ann_light_massless} we obtain the emissivity
\begin{align}
\label{eq:emis:ann:light}
\epsilon_{\rm ann}^{\rm light} =  \epsilon_{\rm max} \left( \frac{T}{30 \, \MeV}  \right)^5 \left( \frac{\sqrt{g_\chi g_\mu}}{3 \times 10^{-5}} \right)^4 \, . 
\end{align}

\subsubsection{Annihilation and scattering contributions to trapping}
\label{sec:ann:trapping}

Given the scattering of a $\chi$ with another particle $b$, the absorption rate can be approximated by (see Appendix~\ref{app:absorption})
\begin{align}
\Gamma_\chi^{b}\approx\left(\prod_i F_{{\rm deg},i}\right)\mathfrak g_b \int\frac{d^3 p_b}{2\pi^3}f_b\sigma(s) v \, ,
\label{eq:GammaAnn}
\end{align}
where $\mathfrak g_b$ are the number of degrees of freedom of the particle $b$ ($\mathfrak g_b=2$ for $b=\bar\chi,\mu,e$ and $\mathfrak g_b=1$ for $b=\nu_\ell$), $v=\sqrt{(p_\chi\cdot p_b)^2- m_\chi^2 m_b^2} /E_\chi E_b$ is the M{\o}ller velocity and $\sigma (s)$ is the scattering cross section for $\chi + b \to X_1 + \hdots + X_n$~\cite{Gondolo:1990dk}. Moreover, the index $i$ runs over the final-state particles and we have approximated the effect of the Pauli blocking by its thermal average or degeneracy factors~\cite{Raffelt:1996wa}
\begin{equation}
\label{eq:fdeg}
F_{\text{deg},i} = \langle  1-f_i \rangle_{i}
= \frac{\mathfrak g_i}{n_i} \int \frac{d^3p_i}{(2\pi)^3} f_i(1- f_i)\, ,
\end{equation}
where $\mathfrak g_i$ and $n_i$ denote their degrees of freedom and number densities, respectively. 

There are two types of processes related  by crossing to the annihilation diagram that are relevant for the trapping regime:  \textit{Inverse annihilation}, $\chi\, \overline{\chi}\to \mu^-\, \mu^+$ and  \textit{scattering}, $\chi\, \mu^-  \to \chi\, \mu^- $ and $\bar\chi\, \mu^-  \to \bar\chi\, \mu^- $. 
Scattering processes are kinematically more involved as they exchange a $Z^\prime$ in the $t$-channel. For very light $Z^\prime$'s ($m_{Z^\prime}\ll T$), they  have a differential cross section with a Coulombian enhancement in the forward direction which involves a small momentum transfer and, therefore, little contribution to the thermalization rate between the dark and SM sectors.   

The interplay between the contributions of inverse annihilation and scattering to the absorption rate is similar to the case of heavy-lepton neutrinos in SN~\cite{Raffelt:2001kv}, as recently emphasized in Ref.~\cite{DeRocco:2019jti}. In the absence of self-interactions between the dark fermions, these two processes really define two surfaces that determine different properties of the dark luminosity in the trapping regime. The freeze-out of inverse annihilation first fixes the number flux of $\chi$'s. The outgoing flow then thermalizes via scattering processes with the leptons until they decouple at a larger radius. For $Z^\prime$ masses in the resonant regime, absorption rates will be dominated by the inverse annihilation and the two surfaces coalesce into a single dark sphere that determines the thermal emission of the $\chi$'s. In the heavy regime both mechanisms can be important and this distinction must be kept in mind.

This is also reminiscent of models with new neutrino self-interactions and their effect in the dynamical evolution of the SN~\cite{Manohar:1987ec,Dicus:1988jh,Bustamante:2020mep,Chang:2022aas,Cerdeno:2023kqo,Fiorillo:2023cas,Fiorillo:2023ytr}. Notably, when considering the model in Eq.~\eqref{eq:LagSimplified} with neutrino interactions, processes such as $\nu_\ell\bar\nu_\ell\to\nu_\ell\bar\nu_\ell$ also occur. These effects may lead to a fundamentally different incarnation of the SN~1987A cooling limit, valid in the trapping regime, which is however beyond the scope of the classical SN~1987A cooling bound that we apply in our analysis.

With all this in mind, our fiducial analysis in the trapping regime includes both processes, inverse annihilation and scattering, in the calculation of the MFP to obtain one single dark sphere that determines $L_\chi$. However, we repeat the calculations for the case where we do not include the scattering processes. We take the variation of our results as an indicator of the potential systematic uncertainties involved in our treatment of the trapping regime. 

In addition, \textit{dark elastic scattering}, $\chi\bar\chi\to \chi\bar\chi$, may become relevant in the trapping regime, as recently discussed in Refs.~\cite{Darme:2020sjf,Sung:2021swd}. In our case, however,  this will not play an important role for the calculations of $L_\chi$ in the trapping regime. The reason is that dark elastic scattering does not directly contribute to maintaining the population of $\chi$'s in thermal equilibrium with the SM plasma after they freeze out (after inverse annihilation turns off). Moreover, for low $\mZ$, the rate of dark elastic scattering is resonant and overwhelmingly larger than scattering or \textit{inverse photoproduction} (discussed below in Sec.~\ref{sec:photoproduction:trapping}), that do tend to maintain the thermal equilibrium between the dark and SM sectors. This situation is particularly relevant for muons, whose population drops significantly at the outer layers of the PNS where freeze out of the $\chi$'s occurs. Therefore, they would effectively decouple immediately after, except for the contributions to the MFP induced by their interactions with the leptons, which are already accounted for in our fiducial analysis\footnote{A proper treatment of the thermalization rates of dark matter with self interactions is, actually, an important issue when studying the structure predicted by these models at the center of galactic halos (see e.g. Ref.~\cite{Tulin:2017ara}).}. Nevertheless, we have studied the contributions of dark elastic scattering processes for completeness, and discuss in Sec.~\ref{sec:Results} the  consequences for our results if these are included  at face value in the calculation of $\Gamma_\chi$ and the MFP. 

\subsection{Photoproduction} 

For muons we use two approximations: \textit{(1)} Describe the Pauli blocking of the final muon by $(1-f_i)\to F_{{\rm deg},i}$ in Eq.~\eqref{eq:fdeg} and, \textit{(2)} in the phase space integrals of the initial particles we take the extreme non-relativistic limit where the muon is static and recoilless. Thus, the kinematics is evaluated in the muon's rest frame, $s=m_\mu^2+2m_\mu \omega$, with $\omega$ the photon's energy,  and $E_\chi+E_{\bar\chi}=\omega$. One arrives at the conventional formula~\cite{Raffelt:1996wa}
\begin{align}
\label{eq:photoproductionNR}
Q_{\gamma}^{\rm nr}=\frac{n_\mu\,F_{{\rm deg},\mu}}{\pi^2}\int_{\omega_0}^\infty d\omega\omega^3 f_\gamma\,\sigma(s)\, ,
\end{align}
where $\omega_0=(s_0-m_\mu^2)/2m_\mu$ and $s_0$ is the kinematic threshold of the process. 

While Eq.~\eqref{eq:photoproductionNR} is approximately valid for muons, electrons are instead ultra-relativistic in the PNS. If we neglect the electron's mass in the phase space integrals\footnote{Since the photoproduction cross sections have a kinematic singularity in the limit $m_e\to 0$, we keep the physical value of the mass in numerical estimates, neglecting also plasma effects.} and assume that $E_\chi+E_{\bar\chi}\approx(\omega + \omega^\prime)/2$, then
\begin{align}
\label{eq:photoproductionR}
Q_{\gamma}^{\rm r}&=\frac{F_{{\rm deg},e}}{8\pi^4}\int^\infty_0 d\omega \omega f_\gamma \int^\infty_0d\omega^\prime\omega^\prime (\omega+\omega^\prime)f_e\nonumber\\
&\times\int^{+1}_{-1}d(\cos\theta)s\sigma(s)\, ,
\end{align}
where $\omega^{\prime}$ is the electron energy and $s=2\omega\omega^\prime(1-\cos\theta)$.      

Similarly to the annihilation topologies, one can estimate the average energy available for the production of dark particles in the photoproduction process. For this we use the average CM energy at threshold,
\begin{align}
\label{eq:AvEnPhotop}
\mathcal E_\ell^\gamma = \sqrt{\langle s\rangle_{\ell\bar\ell}} - m_\ell,
\end{align}
and one can also define different regimes of $\mZ$ for photoproduction: \textit{(1)} the \textbf{``heavy regime''} where $\mZ\gtrsim1\;\GeV \gg \mathcal E_\ell^\gamma$;  \textit{(2)} the {\bf ``resonant regime''} with $m_{Z^\prime} \lesssim \mathcal E_\ell^\gamma$, where the $Z^\prime$ can be produced on-shell. For the conditions specified in Eq.~\eqref{eq:typicalPNS} we find $\mathcal E_\mu^\gamma \sim 90 \, \MeV$ and $\mathcal E_e^\gamma \sim 150 \, \MeV$. 

In Appendix~\ref{app:2to3} we provide the results for the photoproduction cross sections along with some details of the calculations. In addition, we have checked the validity of the approximations in Eqs.~\eqref{eq:photoproductionNR} and~\eqref{eq:photoproductionR} by calculating the \textit{full} 5-body phase space thermal integrals exactly, as described in the seminal work of Ref.~\cite{Petrosian:1967alk}, where this was done for effective (axial)vector operators. For completeness, we outline this calculation in Appendix~\ref{app:full_photoproduction}. We find that the approximate formulas provide a very accurate description, within $\sim$20\% (muons) and $\sim40\%$ (electrons) of the full photoproduction rates in the resonant regime and neglecting degeneracy of the final state lepton. In the heavy regime, Eq.~\eqref{eq:photoproductionNR} underestimates the full  result for muons by a factor $\sim8$. Nevertheless, as shown below in Sec.~\ref{sec:photoproduction:heavy}, the emission rate for heavy $\mZ$ is dominated by annihilation by orders of magnitude. On the other hand, for the highly degenerate electrons in the PNS, we find that approximating the Pauli blocking effects by $F_{\rm deg,e}$ overestimates the full energy loss rate by a factor $\sim3$. In the case of muons, the degeneracy factor $F_{\rm deg,\mu}$ instead agree with the exact result within $\sim5\%$ accuracy.   

In the following, we discuss in more detail the two regimes of photoproduction and the contribution of inverse photoproduction to the interaction rate $\Gamma_\chi$.

\subsubsection{Heavy regime}
\label{sec:photoproduction:heavy}

The cross section of the process $\mu^-\gamma\to\mu^-\chi\bar\chi$, induced by the exchange of a heavy $Z^\prime$, is
\begin{align}
\label{eq:photoproduction-heavy}
\sigma(s)&= \frac{\alpha g_\mu^2 g_\chi^2 m_\mu^2}{1728\pi^2 m_{Z^\prime}^4} \frac{1}{\hat s^2 (\hat s-1)^3}\\
&\times\Big(-55\hat s^6+682\hat s^5 + 483\hat s^4-968\hat s^3 -169\hat s^2 +30\hat s 
\nonumber\\
&-3+12\hat s^2\big( 2\hat s^4 -14\hat s^3 -87\hat s^2 -52\hat s+1\big) \log\hat s\Big)\, ,\nonumber
\end{align}
with $\hat s=s/m_\mu^2$, $\alpha$ denotes the fine-structure constant and we have taken for simplicity the massless limit $m_\chi\to0$. This cross section is equivalent to the expression obtained by Dicus in~\cite{Dicus:1972yr} for the photoproduction of a neutrino pair using only vectorial couplings\footnote{In fact, our results agree with Ref.~\cite{Dicus:1972yr} but disagrees with Ref.~\cite{Raffelt:1996wa} which cites Ref.~\cite{Dicus:1972yr} with the opposite sign in the term $\propto 120\hat s/(\hat s-1)^2$ in Eq.~(3.12). We thank Georg Raffelt for confirming this typo. See App.~\ref{app:2to3} for more details.}. One might attempt to obtain a more simplified expression of the rate by taking the non-relativistic limit in Eq.~\eqref{eq:photoproduction-heavy}. However,  the cross section in this case,
\begin{align}
\sigma(\omega)=\frac{2\alpha g_\mu^2 g_\chi^2}{105\pi^2m_\mu^2}\frac{\omega^4}{m_{Z^\prime}^4}\, , 
\label{NRlimit}
\end{align}
grows rapidly with energy and leads to a gross overestimation of the integral in Eq.~\eqref{eq:photoproductionNR} (see also Ref.~\cite{Petrosian:1967alk}). For instance, for the typical PNS conditions used above, one obtains a rate that is larger than the one obtained using the relativistic expression of the cross section by a factor $\sim25$.

Nevertheless we use the non-relativistic approximation together with Eq.~\eqref{eq:photoproductionNR} to get a rough estimate of the emissivity produced by the photoproduction process in this regime, giving
\begin{align}
\label{eq:emis:photoproduction:heavy}
\epsilon_{\gamma}^{\rm heavy}=\epsilon_{\rm max}\left(\frac{Y_\mu}{0.025}\right)\left(\frac{T}{30\;\MeV}\right)^8\left(\sqrt{g_\chi g_\mu}\,\frac{\TeV}{\mZ}\right)^4\, .
\end{align}
Comparing with the equivalent contribution from annihilation in Eq.~\eqref{eq:emiss_ann_heavy}, we conclude that photoproduction gives an emissivity rate that is smaller by a factor $\sim 250$ and thus can be neglected in the heavy regime.    

\subsubsection{Resonant regime}
\label{sec:photoproduction:resonant}

The resonant $\mu^-\gamma\to\mu^-Z^\prime(\to\chi\bar\chi)$ cross section is
\begin{align}
\label{eq:photoproduction-res}
\sigma(s)&=\frac{\pi\alpha\alpha_\chi}{m_\mu^2}\frac{\text{BR}_{\chi}}{\hat s^2(\hat s-1)^3}\\
&\times\Big(\big(\hat s(\hat s(\hat s+7 x^\prime+15)+2 x^\prime-1)-x^\prime+1\big)R(x^\prime,\hat s)\nonumber\\
&+4\hat s^2\big(\hat s^2-2\hat s(x^\prime+3)+2x^\prime(x^\prime+1)-3\big)\nonumber\\
&\times\tanh^{-1}\frac{R(x^\prime,\hat s)}{\hat s-x^\prime+1}\Big)\nonumber\, , 
\end{align}
where we have introduced the notation $\alpha_\chi=g_\mu^2/4\pi$, and where $\hat s=s/m_\mu^2$, $x^\prime=m_{Z^\prime}^2/m_\mu^2$ and $R(x,\hat s) = (\hat s^2 -2\hat s (x+1) +(x-1)^2)^{1/2}$. For $\text{BR}_{\chi}\to1$ we recover the cross section for semi-Compton production of massive vector bosons, $\gamma\mu^-\to Z^\prime\mu^-$.  In the $\mZ\to0$ limit, one obtains
\begin{align}
\sigma(s)&=\frac{\pi\alpha\alpha_\chi}{m_\mu^2}\frac{\text{BR}_{\chi}}{\hat s^2(\hat s-1)^3}\;\Big(\hat{s}^4+14 \hat{s}^3-16 \hat{s}^2+2 \hat{s}-1\nonumber\\
&+\left(2 \hat{s}^4-12 \hat{s}^3-6 \hat{s}^2\right) \log \hat{s}\Big)\, , 
\end{align}
and if we now perform the non-relativistic expansion,
\begin{align}
\sigma(s)\approx\frac{8\pi\alpha\alpha_\chi\text{BR}_{\chi}}{3m_\mu^2}\, ,
\end{align}
we recover the Thomson cross section for $\alpha_{\chi}\to\alpha$ and $\text{BR}_\chi\to1$. This is the expression commonly used for the semi-Compton production of vector particles in stellar plasmas~\cite{Raffelt:1996wa,Croon:2020lrf}, but it is less appropriate for the PNS where leptons are relativistic. In fact, in case of muons for $\mZ=0$ and in the typical conditions we have been using for the PNS, we find that the Thomson cross section overestimates the relativistic one by a factor $\sim2$. 

On the other hand, the energy-loss rate of the full resonant photoproduction cross section is insensitive to $\mZ$ up to $\mZ\gtrsim~T$, at which point it starts dropping due to increased Boltzmann suppression and defines the onset of the heavy regime. Taking as a reference the Thomson cross section, we can estimate the emissivity of the photoproduction in the resonant regime as
\begin{align}
\label{eq:emis:photoproduction:res}
\epsilon_{\gamma}^{\rm res}=\epsilon_{\rm max}\left(\frac{Y_\mu}{0.025}\right)\left(\frac{T}{30\;\MeV}\right)^4\left(\frac{g_\mu}{5\times 10^{-10}}\right)^2\, ,
\end{align}
using our typical PNS conditions in Eq.~\eqref{eq:typicalPNS}. Comparing this to the emissivity from $\mu^+\mu^-$ annihilation, Eq.~\eqref{eq:emis:ann:light}, we observe that the rate of $\chi\bar\chi$ production from muons for light $Z^\prime$ will be dominated by photoproduction for many orders of magnitude. For $\mZ\lesssim 10$ MeV this process is even more important than resonant neutrino-antineutrino annihilation for the case $g_{\nu_\ell}=g_\mu$, as demonstrated by comparing to Eq.~\eqref{eq:emis:ann:res}.

\subsubsection{Contribution of inverse photoproduction to trapping}
\label{sec:photoproduction:trapping}

Assuming thermal equilibrium, which is adequate in the trapping regime, the contribution of inverse photoproduction $\ell^-\chi\bar\chi\to\ell^-\gamma$ to $\Gamma_\chi$ can be related to the production rates by means of detailed balance (see Appendix~\ref{app:absorption}). In case of muons, the photoproduction rate of $\chi$ per unit volume can be calculated using the same approximations as in Eq.~\eqref{eq:photoproductionNR},  
\begin{align}
\label{eq:photoproductionColl}
\mathcal C_{\rm prod,\mu}^{\gamma}=\frac{n_\mu\,F_{{\rm deg},\mu}}{\pi^2}\int_{\omega_0}^\infty d\omega\omega^2 f_\gamma\,\sigma(s)\, ,
\end{align}
while for electrons we use
\begin{align}
\mathcal C_{\rm prod,e}^{\gamma}&=\frac{F_{{\rm deg},e}}{8\pi^4}\int^\infty_0 d\omega \omega f_\gamma \int^\infty_0d\omega^\prime\omega^\prime f_e\nonumber\\
&\times\int^{+1}_{-1}d(\cos\theta)s\sigma(s)\, .
\end{align}

Then, we estimate the contribution of photoproduction to the MFP by applying detailed balance, using the inverse of $\langle \Gamma_\chi^{\gamma}\rangle$ (see Appendix~\ref{app:absorption})
\begin{align}
\langle \Gamma_\chi^{\gamma}\rangle_\chi = \frac{\mathcal C_{\rm prod}^{\gamma}}{n_\chi}\, .
\end{align}

Note that this approximation differs from the direct calculation of the contribution of inverse annihilation and scattering in Eq.~\eqref{eq:GammaAnn}. In order to combine the two contributions in our estimate of the MFP we use the approximate formula
\begin{align}
\label{eq:The_elephant_formula}
\lambda = \langle \frac{v_\chi}{\Gamma_\chi^{\bar\chi} + \Gamma_\chi^l + \Gamma_\chi^\gamma}\rangle \approx\frac{1}{\langle v_\chi/(\Gamma_\chi^{\bar\chi}+\Gamma_\chi^{l})\rangle^{-1}+\langle \Gamma_\chi^\gamma\rangle/\langle v_\chi\rangle}\, , 
\end{align}
where all the thermal averages are understood to be taken with respect to the $\chi$ kinematics (see Appendix~\ref{app:absorption}). 

\subsection{Other processes and neglected plasma effects}
\label{sec:uncertainties}

In our analysis, we have selected the processes that are dominant for the production and absorption of $\chi$'s in muons and have neglected plasma effects which are expected to be small in this case. 
Electrons in the PNS, on the other hand, are highly degenerate. 
Moreover, in the plasma the electron and photon dispersion relations are significantly modified. The electron mass effectively increases while the photon acquires a longitudinal mode and an effective mass that could enable the decay $\tilde\gamma\to\chi\bar\chi$, where the ``plasmon'' $\tilde \gamma$ includes these collective plasma modes~\cite{1963PhRv..129.1383A,1992ApJ...392...70B,1994ApJ...425..222H,Raffelt:1996wa}.

Nonetheless, the production of $\chi$'s in the heavy regime and neglecting the $\chi$ mass would be analogous to the SM pair-production of heavy-lepton neutrinos from the electrons in the stellar plasma. Plasmon decay is indeed an important process in the conditions of high densities predicted in the PNS. However, at the high temperatures reached in the SN explosions considered in this work, $e^+e^-$ annihilation becomes the dominant process~\cite{1994ApJ...425..222H}. Adding mass to the $\chi$'s will not affect this conclusion and may, in fact, kinematically close the plasmon decay if $m_\chi\gtrsim 10~\MeV$, which is the scale of the plasma frequencies expected in the PNS. Finally, accounting for the increase of the electron mass in the plasma by a similar amount does not affect significantly the annihilation rates as discussed in Sec.~\ref{sec:AnnHeavy}. 

On the other hand, for the light $Z^\prime$ some of the neglected effects can become important. For instance, resonant  bremsstrahlung production $e^-p\to e^-p (Z^\prime\to\chi\bar\chi)$ could become the dominant process for $\mZ\sim 10~\MeV$, as it is the case for on-shell production of axion-like particles~\cite{Lucente:2022vuo}. In addition, for these masses one needs to consider medium-induced $\tilde\gamma-Z^\prime$ mixing which may also have an impact~\cite{An:2013yfc}. For all of these reasons, our results for the emission from electrons in the light regime, $\mZ\lesssim 50$ MeV, should be considered as an intermediate step towards a more refined and robust calculation, and the results that we report for this case regarded as a rough approximation. 

\section{Results} 
\label{sec:Results}

In this section, we apply the upper limit on the dark luminosity in Eq.~\eqref{eq:Lchi} to obtain the SN~1987A constraint on the parameter space of the different dark sector models we consider. We start by presenting an analysis in terms of effective operators, which corresponds to the heavy regime introduced in Sec.~\ref{sec:SN} for the calculation of the relevant processes. This allows us to generalize our analysis to generic interactions of dark fermions coupled to leptons via dimension-6 operators, and extract a SN~1987A limit for any portal mediator with mass much larger than the temperatures and chemical potentials in the PNS. It is interesting to note that this approach was already applied in the context of neutrino emission from stellar plasmas in the very early days of the SM~\cite{1976PhRvD..13.2700S,1977PhRvD..15..977D}. We then focus on the constraints on the parameter space of the simplified $Z^\prime$ model of Eq.~\eqref{eq:LagSimplified}, which can be regarded as the continuous extension of the EFT bounds to low mediator masses for one particular operator ($VV$). Finally, we present the SN~1987A constraints for a phenomenologically relevant and UV-motivated version of the $Z^\prime$ model, obtained by gauging the $L_\mu-L_\tau$ symmetry. 

\subsection{Effective field theory}
\label{sec:EFT}

\begin{table}[t]
\renewcommand{\arraystretch}{1.8}
  \setlength{\arrayrulewidth}{.25mm}
  \setlength{\tabcolsep}{0.5 em}
\centering
\begin{tabular}{|c|c|c|c|c|}
\hline 
$X Y$ & $\Lambda_\mu^{\rm eff}$ [TeV] & $\Lambda_{\nu_\mu}^{\rm eff}$ [TeV] & $\Lambda_e^{\rm eff}$ [TeV] \\
\hline 
$SS$ & $0.0017 - 4.4$ & $0.062 - 5.2$ & $0.070 - 5.4$\\
$PS$ & $0.00044 - 5.1$ & $0.062 - 5.2$& $0.070 - 5.4$\\
$VV$ & $0.0017 - 5.7$ & $0.072 - 5.6$ & $0.11 - 5.9$\\
$AV$ & $0.0022 - 4.7$ & $0.072 - 5.6$ & $0.11 - 5.8$ \\
$LL$ & $0.0015 - 3.7$ & $0.051 - 4.0$ & $0.074 - 4.1$ \\
$LV$ & $0.0018 - 4.4$ & $0.061 - 4.7$ & $0.088 - 4.9$ \\
$TT$ & $0.0033 - 6.8$ & $0.10 - 6.7$   & $0.17 - 7.0$ \\
\hline
\end{tabular}
\caption{SN~1987A exclusion range of the effective scale $\Lambda_l^{\rm eff} \equiv \Lambda_l/\sqrt{C_{X Y}^l}$ for the EFT interactions defined in Eq.~\eqref{eq:EFTLag}, with $m_\chi = 0$ and for the simulation SFHo-18.8. The lower limit of the constraint is set by the trapping regime. The bounds for $XY=SP$, $PP$, $AA$, $VA$, and $T^\prime T$ are equal to those of $SS$, $PS$, $VV$, $AV$, and $TT$, respectively. The bounds for $RR$, $LR$ and $RL$ are  identical to those of the $LL$ operators. The constraints on $\Lambda^{\rm eff}_{\nu_e}$ essentially coincide with those on $\Lambda^{\rm eff}_{\nu_\mu}$.
\label{Tab}}
\end{table}

\begin{figure}[tb]
\centering 
\includegraphics[width=1.0\columnwidth]{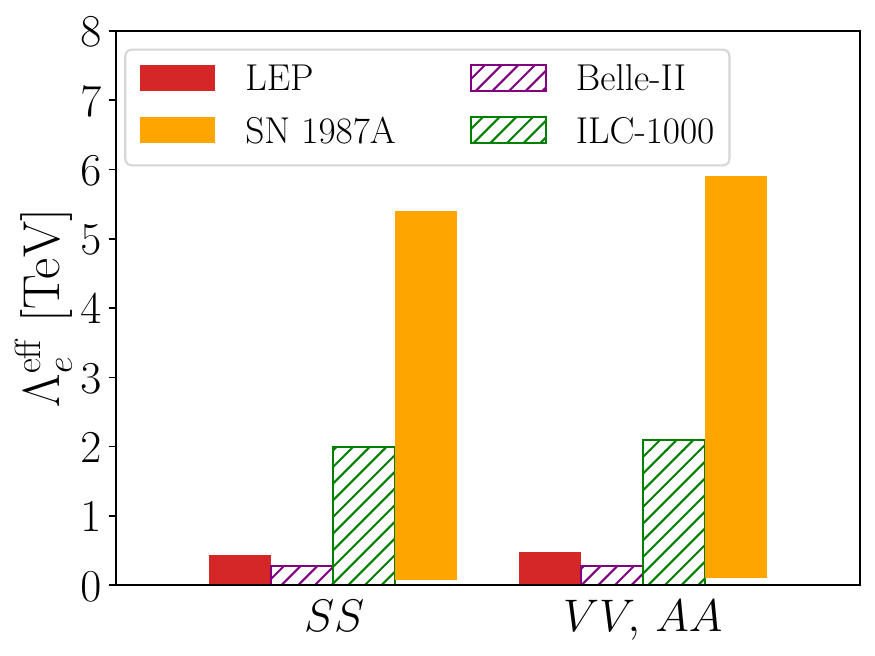}
\caption{ \label{fig:EFTbounds} Comparison between SN~1987A and collider limits on the effective scale $\Lambda_e^{\rm eff} \equiv \Lambda_e/\sqrt{C_{X Y}^e}$ for the EFT interactions of electrons defined in Eq.~\eqref{eq:EFTLag}, with $m_\chi = 0$. The orange bars show the exclusion range obtained from SN~1987A, using the simulation SFHo-18.8. The upper limit is obtained by the free-streaming regime and the lower limit is set by the trapping regime. The red bars show the lower limits obtained at LEP~\cite{Fox:2011fx}, while the purple and green hatched bars show the projections for Belle-II with $50\, {\rm ab}^{-1}$ of integrated luminosity~\cite{Liang:2021kgw} and ILC with $\sqrt{s}=1\, {\rm TeV}$ and $ L_{\rm int} = 1000\, {\rm fb}^{-1}$~\cite{Chae:2012bq}.}
\end{figure}

\begin{figure}[t]
\centering 
\includegraphics[width=1.0\columnwidth]{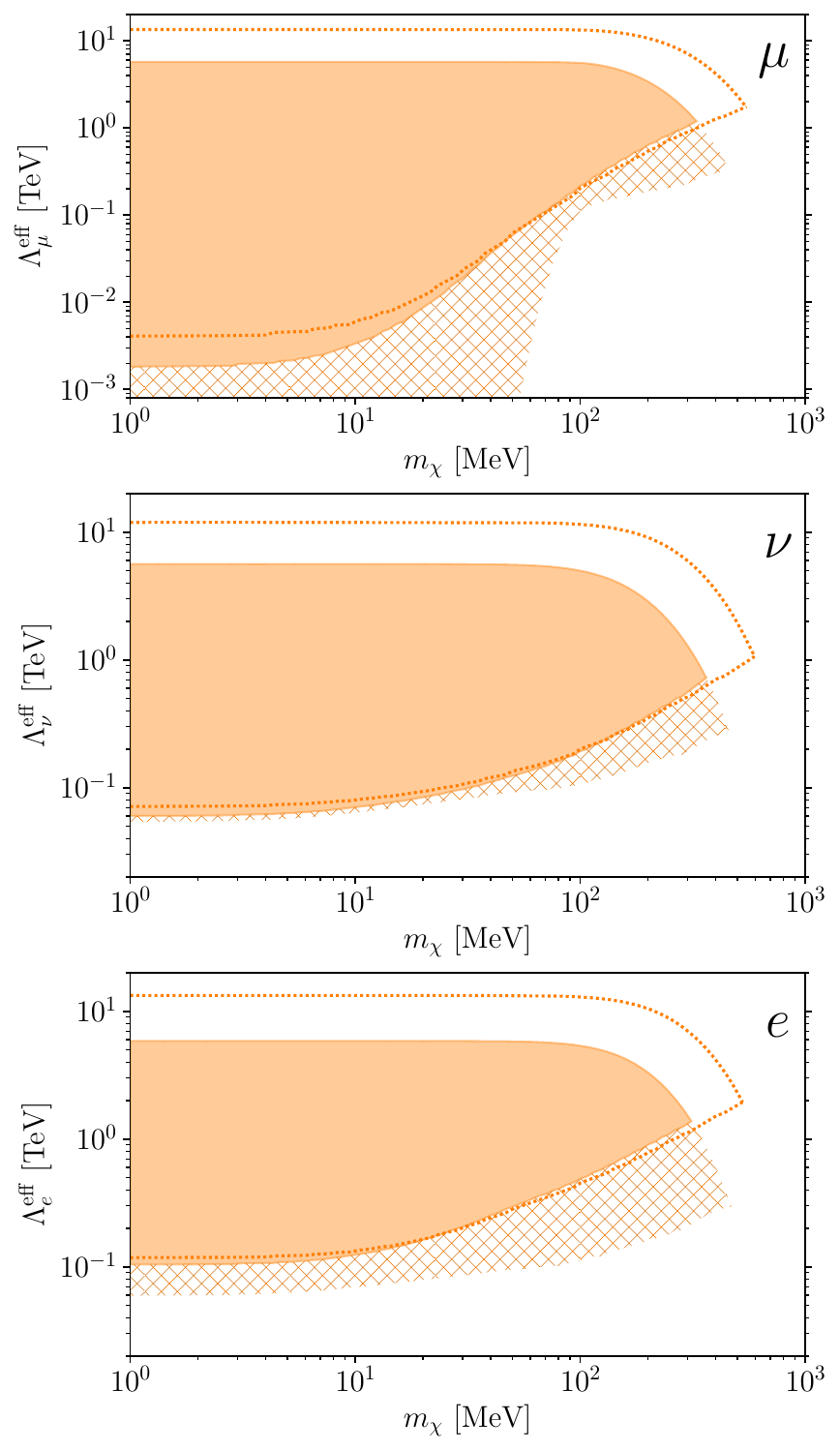}
\caption{\label{eft} SN~1987A constraints on the heavy scale $\Lambda_{l}^{\rm eff}$ as a function of the  dark fermion mass $m_\chi$ for the $VV$ operator and different leptons. The solid orange region is excluded using numerical input from the simulation SFHo-18.8, while the dotted line encloses the region excluded using SFHo-20.0. The hatched region gives an estimate of the uncertainty of the trapping regime, obtained by omitting the contribution to the MFP from scattering processes, as discussed in Sec.~\ref{sec:ann:trapping}. The constraints on $\nu_e$ and $\nu_\mu$ are essentially similar, so we show those on $\nu_\mu$ and label them as $\nu$.}     
\end{figure}

The analysis for the heavy $Z^\prime$ can be generalized in the context of an EFT with four-fermion operators. Focusing on the couplings of dark-sector fermions $\chi$ to leptons $l=e$, $\mu$, $\nu_\ell$, the most general effective Lagrangian at leading order is  
\begin{align}
\label{eq:EFTLag}
{\cal L}_{\rm EFT} &= \frac{1}{\Lambda_l^2}\sum_{X,Y} 
C_{X Y}^l\left(\overline l\, 
\Gamma_X\, l\right)\cdot \left(\overline{\chi}\,\Gamma_{Y}\,\chi \right)\, ,
\end{align}
where $X, Y$ run over $V,A,S,P, L,R, T,T'$, with $\Gamma_V=\gamma^\mu$, 
$\Gamma_A=\gamma^\mu\gamma_5$, $\Gamma_S=\mathbb{1}$, $\Gamma_P=\gamma_5$,  $\Gamma_{R,L}=\gamma^\mu(\mathbb{1} \pm \gamma_5)/2$,  $\Gamma_T=\sigma^{\mu\nu} = i/2 [\gamma^\mu, \gamma^\nu]$  and $\Gamma_{T'}=\sigma^{\mu\nu}\gamma_5$, and Lorentz  
indices properly contracted\footnote{Only two 
tensor operators are independent and we will take those corresponding to $C^l_{TT}$ and 
$C^l_{T'T}$.}. Matching the effective operators to the $Z^\prime$ model in Eq.~\eqref{eq:LagSimplified} coupled to charged leptons yields $\Lambda_\ell = \mZ$ and $C_{VV}^\ell= g_\ell g_\chi$. For neutrinos their bilinears in the EFT Lagrangian are constructed with left-handed fields and contribute, instead, to $C_{LV}^{\nu_\ell}=g_{\nu_\ell}g_\chi$.   

In Table~\ref{Tab}, we show the limits obtained on $\Lambda_l^{\rm eff} \equiv (\Lambda_l/C^l_{X Y})^{1/2}$ for these interactions in the limit $m_\chi=0$ for muons, neutrinos and electrons. The upper limits correspond to the free-streaming regime, the lower limits to the trapping regime. Notice that the excluded regions of $\Lambda_l^{\rm eff}$ are in the EFT range of validity as long as $\Lambda_l\gtrsim 1 \, \GeV$, which is much larger than all other energy scales relevant in the PNS. This is the case for essentially all operators. 

The SN~1987A bounds on the EFT operators are very strong, reaching up to $4-7$ TeV. This sensitivity to the mediator mass scale is approximately one order of magnitude better than the one achieved by laboratory experiments for similar leptonic interactions~\cite{Fox:2011fx,Alcaide:2019pnf,Escrihuela:2021mud,Liang:2021kgw}~\footnote{See also~\cite{Fernandez-Martinez:2023phj} for an estimate of the SN~1987A  bounds obtained by recasting the results derived in~\cite{DeRocco:2019jti} for an operator coupled like the electromagnetic current.}. For instance, monophoton searches at LEP have been used to set the following lower bounds on $\Lambda^{\rm eff}_e$~\cite{Fox:2011fx},
\begin{align}
\begin{split}
VV,~AA:&\hspace{0.6cm}0.48\text{ TeV}\, ,\\
SS:&\hspace{0.6cm}0.44\text{ TeV}\, .
\end{split}
\label{eq:LEP_EFT}
\end{align}

In Fig.~\ref{fig:EFTbounds}, we show the SN~1987A limits on these EFT operators compared with those obtained at LEP~\cite{Fox:2011fx}, and with the projections of the sensitivity that could be achieved at Belle II~\cite{Liang:2021kgw} or at a future $e^+e^-$ linear collider~\cite{Chae:2012bq}. Remarkably, the SN~1987A bounds are stronger than LEP by roughly one order of magnitude, and will even dominate over future collider limits. Note however that SN~1987A constraints apply only to sufficiently light dark fermions, while collider bounds typically extend to larger $\chi$ masses, such as LEP, which provides constraints for $m_\chi\lesssim100$ GeV. 

As discussed in Sec.~\ref{sec:SN}, the dominant effect in the heavy regime of the $Z^\prime$ model is annihilation (for typical SN conditions), and  photoproduction can be neglected. We have checked that this is indeed true for any EFT operator by calculating their contributions to photoproduction explicitly (see Appendix~\ref{app:2to3}). In fact, we note that, for the EFT limit of the $Z^\prime$ model, the approximate expression in the high-energy limit ($m_\ell\to0$) in  Eq.~\eqref{eq:Q_ann_heavy_massless} leads to a bound on the effective scale that is off only by $\sim30\%$ with respect to the full calculation. Therefore, since the high-energy limit  gives quite accurate results,  the numerical bounds for different Lorentz structures and different leptons are of the same order, as in this limit the corresponding cross sections differ at most by ${\mathcal O}(1)$ numbers. 

In Fig.~\ref{eft} we show the dependence of the limits on the heavy scale as a function of the dark fermion mass $m_\chi$ for the effective $VV$ interaction and $LV$ for neutrinos. As expected, the excluded regions shrink with increasing $m_\chi$ and they are limited to masses $m_\chi\lesssim 300$ MeV. We also analyze the variations of the constrained region produced by using the simulation SFHo-20.0 or different prescriptions for the processes included in the trapping regime.  The upper limits of $\Lambda_l^{\rm eff}$ obtained from this hotter simulation are a factor $\sim2$ stronger in the free-streaming regime, because in this case emission is dominated by the hottest region inside the supernova. On the other hand, the uncertainties of the boundary with the trapping regime 
for electrons and neutrinos 
are relatively small and our fiducial calculation is, again, on the conservative side. The reason for this behavior is that the dark sphere is not located in the region of highest temperature.
This makes trapping sensitive to the shape of the temperature profile, which is similar for both simulations.

Moreover, the scale marking the onset of the trapping regime in these cases is $\Lambda_l^{\rm eff}\approx100$ GeV, which is of the order of the electroweak scale. This is consistent with the fact that the boundary of the trapping regime is set by the  luminosity of the trapped neutrinos via Eq.~\eqref{eq:Lchi}, which interact with SM leptons precisely through dimension-6 operators suppressed by the Fermi scale.

For muons, however, the bound extends to scales that are a few orders of magnitude lower than for electrons and neutrinos. In addition, there is a large variation in the location of the  boundary depending on the selection of processes contributing to trapping. The reason is that muons are relatively heavy and there is a maximal radius where they can be produced by thermal fluctuations in the plasma. Putting it differently, there are no muons to scatter with in case of inverse photoproduction and inverse annihilation becomes ineffective because of the strong phase-space suppression. In case that only $\mu-\chi$ interactions are included in the calculation of the MFP, the radius of the dark sphere is typically smaller than the one of the neutrino sphere and the bounds on $\Lambda_\mu^{\rm eff}$ extend down to $\sim 1~\GeV$. However, if one were to also include $\chi\bar\chi$ elastic scattering in the calculation of the MFP, and the $\chi$ is light, $m_\chi\lesssim T$, then they would produce contributions  analogous in size to those given by the neutrino and electron interactions and the boundary of the trapping regime would be increased again to $\sim 100~\GeV$, cf.~the discussion in Section~\ref{sec:ann:trapping}.                                

\subsection{Simplified $Z^\prime$ models}
\label{sec:simplifiedZp}

\begin{figure}[t]
\centering 
\includegraphics[width=1.0\columnwidth]{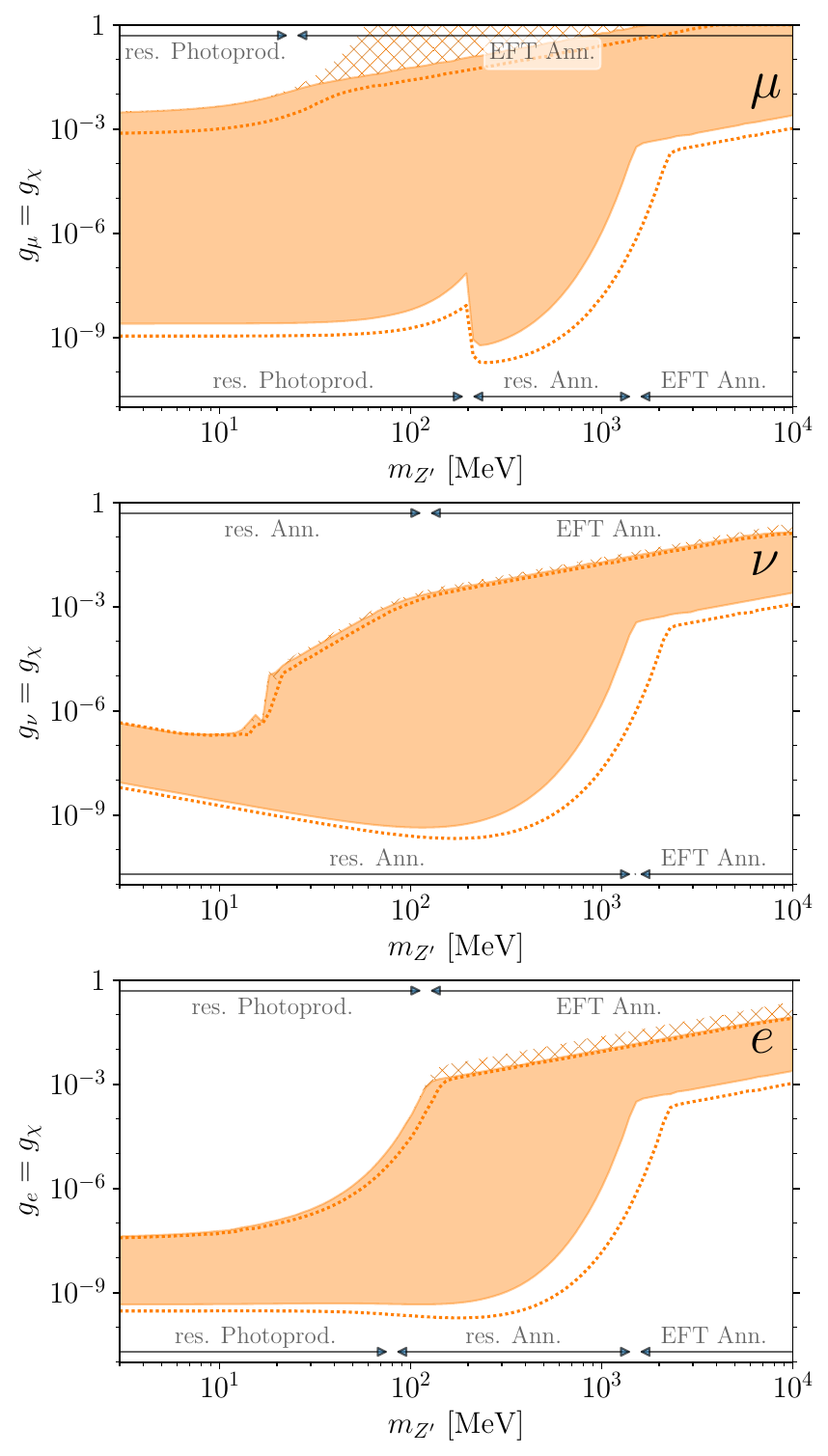}
\caption{ \label{regimes} Excluded parameter space of the $Z^\prime$ model in Eq.~\eqref{eq:LagSimplified} using constraints from SN~1987A with the simulation labeled as SFHo-18.8 in Ref.~\cite{Bollig:2020xdr}. The dotted line encloses the region excluded using SFHo-20.0. 
The hatched region gives an estimate of the uncertainty of the trapping regime, obtained by omitting the contribution to the MFP from elastic scattering processes, as discussed in Sec.~\ref{sec:ann:trapping}.  }
\end{figure}

We now present the SN~1987A constraints in the parameter space of the simplified $Z^\prime$ model in Eq.~\eqref{eq:LagSimplified}. In the derivation of our results for the annihilation contributions to $Q$ we numerically solve the integral in Eq.~\eqref{eq:Qann} (and the equivalent ones for neutrinos and electrons), as described in Appendix~\ref{app:numerical}. This procedure allows us to track the contribution of the annihilation rates across the whole $\mZ$ range, including the transition regions between the three regimes defined in Sec.~\ref{sec:annihilation}. Instead, for photoproduction we only use the expressions in the resonant regime, because photoproduction  is only relevant for the low-$\mZ$ regime of the charged leptons, see Eq.~\eqref{eq:emis:photoproduction:heavy}. 

\begin{figure*}[hbt!]
\centering 
\includegraphics[width=0.95\textwidth]{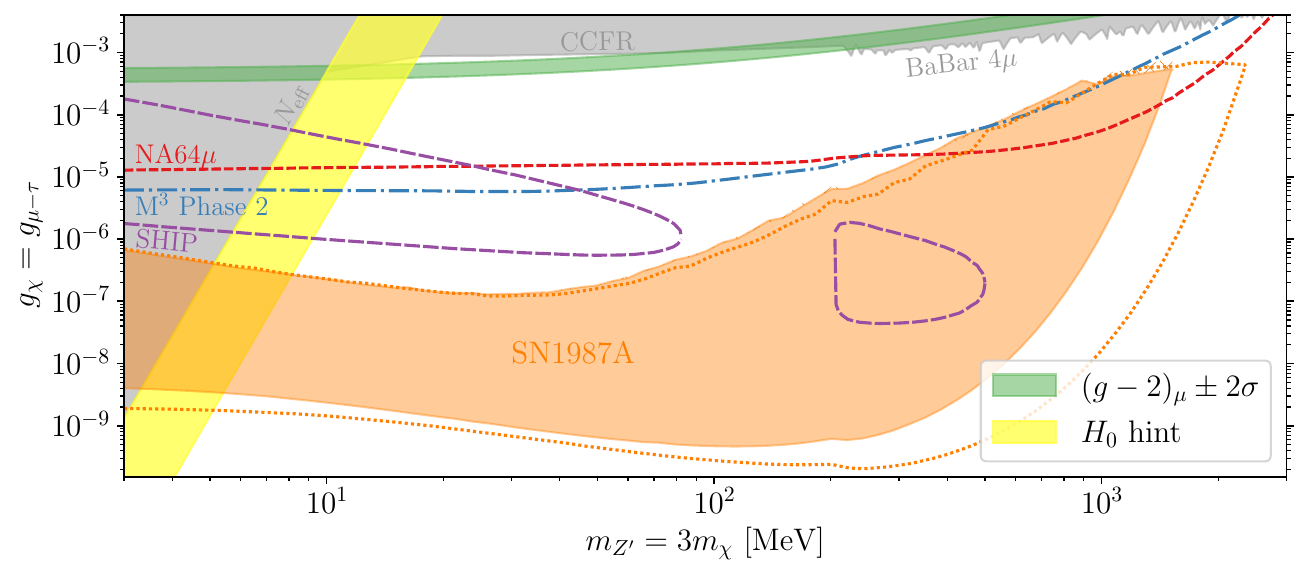}
\caption{SN~1987A cooling constraint (orange region) in the mass-coupling plane of the $L_\mu-L_\tau$ model, assuming $\mZ=3m_\chi$ and $g_\chi=g_{\mu-\tau}$. The dotted orange line indicates the bound obtained from using the SN simulation SFHo-20.0, instead of SFHo-18.8 which gives the weakest constraints.
Also shown are the preferred regions to explain the $(g-2)_\mu$ anomaly (green band) and the $H_0$ tension (yellow band), along with current constraints (gray region) and the forecasts for future experiments. See main text for details. 
\label{third} }
\end{figure*}

In Fig.~\ref{regimes} we show the SN~1987A limits for the model in Eq.~\eqref{eq:LagSimplified}, for the case when only one of the leptonic couplings is present and with $g_\chi=g_l$. In the lower (upper) part of the plots we identify the dominant process for production (absorption) in the indicated mass range. We also show the variation of the constrained region obtained by using the SFHo-20.0 simulation (dotted curve) and, independently, by omitting the scattering contribution to the MFP in the trapping regime (hatched region). In all cases we observe the onset at high masses of the power-law behavior of the constraints $\propto1/\mZ^{4}$, characteristic of the EFT. Interestingly, this occurs for $\mZ\gtrsim1$ GeV in the free-streaming domain but already for $\mZ\gtrsim100$ MeV in the trapping regime. This is because in the former case emission is governed by the conditions in the hottest region of the PNS, while in the latter it corresponds to the cooler outmost layer of the dark sphere, where all relevant energy scales are smaller. The boundary of the trapping regime changes very little with respect to adding or not scattering contributions in the absorption rate. Adding dark elastic scattering ($\chi\bar\chi\to\chi\bar\chi$) in the muonic case (for $m_\chi\lesssim T$) would instead  make the trapping region similar to the neutrino case.   

The shape of the constrained region below $\mZ\approx1$ GeV depends on the lepton considered. For muons the low $\mZ$ region is dominated by (resonant) photoproduction, which gives a flat bound up to $\mZ\gtrsim T$, where the on-shell production of the $Z^\prime$ starts decreasing due to Boltzmann suppression. Nevertheless, it remains more important than $\mu^+\mu^-$-annihilation (in the light regime), until it becomes resonant at the $\mu^+\mu^-$ threshold,  $\mZ\geq2m_\mu$. However, this occurs already at large energies and the production quickly suffers from the Boltzmannian suppression, converging to the EFT scaling at higher $\mZ$.  

For electrons, photoproduction dominates again for low $\mZ$, even though the $e^+e^-$ threshold is much lower than for muons.  Resonant annihilation of electrons quickly replaces photoproduction as the dominant process above $\mZ\gtrsim T$ in the free-streaming domain, until the resonance starts suffering from Boltzmann suppression and the EFT takes over. For the trapping regime, there is no range of $\mZ$ where resonant annihilation dominates and the EFT directly replaces inverse-photoproduction at $\mZ\gtrsim 100$ MeV. 

For neutrinos there is no photoproduction and the production and absorption rates are given by annihilation in the resonant and heavy regimes. In the free-streaming regime we observe a strengthening of the bound up to $T\lesssim100$ MeV. This is due to the $\mZ^2$ scaling of the emission rate in the resonant regime, see Eq.~\eqref{eq:Q_ann_res_massless}, which is quickly overcome by Boltzmann suppression until the heavy regime takes over. 

\subsection{Gauged $L_\mu - L_\tau$ models}
\label{sec:gaugedLmuLtau}

Finally, we study SN~1987A constraints on a UV-motivated realization of the simplified $Z^\prime$ model in Eq.~\eqref{eq:LagSimplified}. This is the gauged $L_\mu - L_\tau$ model coupled to  DM fermions, described by the interaction Lagrangian
\begin{align}
{\cal L}_{\rm int} & = Z_\mu^\prime \left( g_{\mu - \tau} j^\mu_{\rm SM} + g_\chi \overline{\chi} \gamma^\mu \chi \right) \, ,    \label{eq:LmumLtauModel}
\end{align}
where $\chi$ is the dark fermion and $j^\mu_{\rm SM}$ is the SM part of the $L_\mu - L_\tau$ current. These interactions induce an irreducible contribution to the kinetic mixing of the $Z^\prime$ with the photon, through $\mu$ and $\tau$ loops, giving a mixing parameter of the order $\epsilon \sim g_{\mu-\tau}/70$~\cite{Kamada:2015era}. This would presumably give only small corrections to our analysis (see e.g.~\cite{Croon:2020lrf}) and thus will be neglected.

It is well known that such models can accommodate the present $(g-2)_\mu$ anomaly with couplings of order $g_{\mu - \tau} \sim 10^{-4}$ for a  light $Z^\prime$ gauge boson, $m_{Z^\prime} \ll m_\mu$~\cite{Fayet:2007ua,Pospelov:2008zw, Chen:2017awl}. They also allow to reproduce the DM relic abundance through resonant $s$-channel annihilation, when muon and  DM fermion have similar couplings to the $Z^\prime$ gauge boson and the latter is heavier than the DM fermion by a factor $2-3$~\cite{Holst:2021lzm, Drees:2021rsg}. For such light masses, $Z^\prime$ decays and DM annihilation can heat the SM bath after neutrino decoupling, thereby increasing the effective number of relativistic degrees of freedom, usually expressed as an effective number of neutrino species $N_{\rm eff}$. Such a contribution could help to reduce the long-standing tension between local and cosmological determinations of the Hubble constant~\cite{Bernal:2016gxb,DiValentino:2021izs}, if the new contributions is of order $\Delta  N_{\rm eff} \sim 0.1-0.4$~\cite{Planck:2018vyg}. 

At present, the most relevant laboratory constraints (see Ref.~\cite{Bauer:2018onh} for an overview) stem from BaBar searches for $Z^\prime$-bosons above 212 MeV decaying to muons~\cite{BaBar:2016sci}, neutrino trident production~\cite{Altmannshofer:2014pba} at CCFR~\cite{CCFR:1991lpl},  bounds on coherent elastic neutrino nucleus scattering from the COHERENT collaboration~\cite{Cadeddu:2020nbr} and constraints on neutrino-electron scattering~\cite{Harnik:2012ni, Kaneta:2016uyt} at BOREXINO~\cite{Bellini:2011rx}. A variety of accelerator-based  searches have been proposed to explore the unconstrained parameter space, such as NA62~\cite{Krnjaic:2019rsv}, which looks for final state radiation of  $Z^\prime$-bosons in $K^+ \to \mu^+ \nu_\mu$, and dedicated searches using muon beam facilities such as   the NA64$\mu$ experiment~\cite{Gninenko:2020hbd} at CERN and $M^3$~\cite{Kahn:2018cqs} at Fermilab.

Astrophysical limits from white dwarfs have been studied already in Refs.~\cite{Dreiner:2013tja,Bauer:2018onh}. Here we show that also constraints from SN~1987A disfavor a large region of  parameter space with significant overlap with the expected reach of planned experiments and with the region that could address the $H_0$ tension. In Fig.~\ref{third}, we show the SN~1987A limits on the muon coupling as a function of the $Z^{\prime}$ mass, in the scenario where muons and dark matter couple equally to the $Z^{\prime}$, $g_{\mu - \tau} = g_\chi$, and $m_{Z^{\prime}} = 3 m_{\chi}$. This is compared to the current bounds discussed above, shown in gray, and the regions preferred at $95\%$ C.L. by $(g-2)_{\mu}$ and $H_{0}$. 

In the free-streaming regime we use the same method as for the simplified model described in Appendix~\ref{app:numerical}, including the contributions from $\mu^+\mu^-$, $\nu_\mu \overline{\nu}_\mu$ and $\nu_\tau \overline{\nu}_\tau$ annihilations. We also add the resonant photoproduction off muons which dominates the rate up to $\mZ\lesssim10$ MeV, where neutrino annihilation starts to give the largest contribution to the rate, producing the characteristic strengthening of the bound with $\mZ$.  At $m_{Z'} = 2\, m_{\mu}$ a small feature signalizes the onset of resonant $\mu^+\mu^-$ annihilation. Instead, the boundary of the constrained region in the deep trapping regime is dominated by inverse resonant annihilation into neutrinos. Note that the $\chi$ mass scales with the $Z^\prime$ mass, which has the effect of slightly suppressing the rate as compared to the massless-$\chi$ case, cf.~Fig.~\ref{regimes}. A rough estimate of the uncertainty of the excluded region is indicated with a dashed orange line, which shows the limits obtained from employing the hottest SN simulation SFHo-20.0, as described in Sec.~\ref{sec:SN}.

\section{Summary and Conclusions}
\label{sec:Conclusions}

In this paper, we have studied the SN~1987A cooling constraints on dark-sector models induced by the emission of new light dark fermions $\chi$ coupled to leptons. To provide a concrete framework, we consider general vector portal interactions arising from the exchange of a massive $Z^\prime$ vector mediator. We focus primarily on the couplings to muons, which are predicted to have sizable number densities within the hot and dense environments of the proto-neutron star formed during core-collapse supernovae. However, we also extend our analysis to couplings to neutrinos and electrons. 

We have considered various mechanisms for the production and absorption of the $\chi$'s and different regimes that depend on the ranges of the parameters of the model. Firstly, the constraints depend on the mass of $\chi$, as their pair production becomes Boltzmann suppressed for $m_\chi\gg T$. Secondly, different regions of $\mZ$ can be identified based on whether the dark fermions are resonantly produced or generated from the tail of the $Z^\prime$ resonance. This distinction arises, for instance, when the $Z^\prime$ is heavy and cannot be produced on-shell by the thermal fluctuations in the medium. Finally, there exist two distinct regimes of coupling values, depending on whether the dark fermions free-stream or become trapped within the PNS. Consequently, by analyzing the $\chi\bar \chi$ production  within these two regimes, for given masses of the $Z^\prime$ and the $\chi$, we can determine the range of couplings that is excluded by the observations from SN~1987A.

For $Z^\prime$ particles with masses $\mZ\lesssim T\sim10$ MeV and massless $\chi$, the observations from SN~1987A place constraints on the couplings between $\sim10^{-1}$ and $\sim10^{-9}$, for equal couplings of the vector mediator to leptons and dark fermions, cf. Fig.~\ref{regimes}. However, the range of these bounds strongly depends on the  $Z^\prime$ mass and the specific lepton to which the $Z^\prime$ couples.

These calculations can be readily extended to explicit  $Z^\prime$-models, for example motivated by a gauged lepton flavor symmetry with a dark sector charged under it. We specifically investigate the case of $L_\mu-L_\tau$, which has been proposed in the literature as a combined solution to the $(g-2)_\mu$ anomaly, the Hubble tension and the dark matter puzzle. The SN~1987A limit covers a large region of  parameter space that overlaps with the forecasts of future experiments and with part of the region that could address some of the tensions, see Fig.~\ref{third}. 

On the other hand, when the $Z^\prime$ mass is larger than the temperature and chemical potentials in the PNS, the interactions mediated by the $Z^\prime$ can be accurately described by  effective operators. This allows us to generalize the analysis to completely generic heavy portal interactions between dark fermions and SM leptons, summarized in Fig.~\ref{eft}. We find that SN~1987A cooling can probe new-physics scales up to $4-7$ TeV (cf.~Table~\ref{Tab}), which surpasses current bounds from laboratory experiments by an order of magnitude (see Fig.~\ref{fig:EFTbounds}).

We emphasize that our analysis is not complete when a light $Z^\prime$  is coupled to electrons. In this case, bremsstrahlung processes are expected to provide the dominant contributions to the emission of dark fermions, and plasma effects can have a significant impact on the analysis. Nevertheless, we consider our results as an important step into this direction, which significantly extends previous studies in the literature. 

Finally, other aspects of SN physics could lead to  constraints complementary to the ones obtained in this work (see e.g. Refs.~\cite{Caputo:2022mah,Fiorillo:2022cdq}). In particular, there have been recent efforts towards a better understanding of the effect of neutrino self-interactions~\cite{Chang:2022aas,Fiorillo:2023cas,Fiorillo:2023ytr} in the trapping regime. Moreover, it would be interesting to check whether bounds obtained from energy transport by dark particles exceed the bounds derived from energy loss alone, following the analysis in Ref.~\cite{Caputo:2022mah}.

\section*{Code availability}
We provide a minimal code example to test different parameter points of the $L_\mu-L_\tau$ model at \url{https://github.com/spinjo/SNforMuTau.git}.

\section*{Acknowledgments}
We thank Andrea Caputo, Pilar Coloma, Miguel Escudero, Patrick Foldenauer, Felix Kahlh\"ofer, Enrico Nardi, Uli Nierste, Filippo Sala and Stefan Vogl for useful discussions. We want to thank Robert Bollig and Hans Thomas Janka for sharing data of the simulations with us. This work is partially supported by project C3b of the DFG-funded Collaborative Research Center TRR257, ``Particle Physics Phenomenology after the Higgs Discovery" and has received support from the European Union's Horizon 2020 research and innovation programme under the Marie Sk\l{}odowska -Curie grant agreement No 860881-HIDDeN. The work of C.A.M. is supported by the Office of High Energy Physics of the U.S. Department of Energy under contract DE- AC02-05CH11231. Work by JMC is supported by PGC2018-102016-A-I00, and the ``Ram\'on y Cajal'' program RYC-2016-20672. 
\appendix
\section{Generalities of absorption rates}
 \label{app:absorption}
In the case of generic $2\to n$ processes, where a dark particle $\chi$ interacts with a particle $b$ in the initial state, the resulting contributions $\Gamma_\chi^{b}$ to the absorption rate of $\chi$ can be approximated by
\begin{align}
\Gamma_\chi^{b} & \approx  \frac{1}{2 \mathfrak g_\chi  E_\chi}\prod_{\text{final }i}F_{\text{deg,}i}\int \frac{d^3p_b}{(2\pi)^3 2E_b} f_b \frac{d^3p_i}{(2\pi)^3 2E_i}\notag\\
&\times(2\pi)^4 \delta^4 (\sum_i p_i) \sum_{\rm spins} |\mathcal{M}|^2 
\notag \\
& =\prod_{\text{final }i}F_{\text{deg,}i} \times n_b\langle\sigma v\rangle_b\, ,
\label{eq:gammaa}
\end{align}
where $\langle \cdot\rangle_b$ denotes the thermal average taken over the particle $b$,  defined by 
\begin{align}
\langle X \rangle_b & = \frac{\mathfrak{g_b}}{n_b} \int \frac{d^3p_b}{(2\pi)^3} f_b X \, . 
\end{align}
Furthermore, $\sigma$ denotes the cross section of the process, $v= \sqrt{(p_b \cdot p_\chi)^2 - m_\chi^2 m_b^2 }/E_b E_\chi $ is the M\o{}ller velocity and we have approximated the Pauli-blocking effects by introducing the degeneracy factors defined in Eq.~\eqref{eq:fdeg}. 

By performing the $d^3p_\chi$ integral in Eq.~\eqref{eq:gamma_chi}, one arrives at
\begin{equation}
\mathcal C_{\rm abs}^b = n_\chi \langle \Gamma_\chi^{b}\rangle_\chi = n_\chi n_b \langle \sigma v\rangle_{\chi\,b} \, ,
\label{eq:avMFP}
\end{equation}
where $\langle \sigma v\rangle_{\chi\,b}$ denotes the thermal average over the complete initial state kinematics, i.e. 
\begin{align}
\langle \sigma v\rangle_{\chi\,b} \equiv \frac{\mathfrak g_\chi \mathfrak g_b}{n_\chi  n_b } \int \frac{d^3 p_{\chi}}{(2 \pi)^3} f_\chi \frac{d^3 p_{b}}{(2 \pi)^3} f_b \sigma (p_\chi, p_b) v  \,.
\label{eq:thermalaverage2}
\end{align}
The inverse process defines an analogous collision operator for production of $\chi$'s, $\mathcal C_{\rm prod}^b$, which in conditions of thermal and chemical equilibrium reads
\begin{align}
\mathcal C_{\rm prod}^b=\mathcal C_{\rm abs}^b\, .
\label{eq:detailed_balance}
\end{align}
This detailed balance relation can then be used to estimate the MFP in the PNS 
\begin{align}
\langle \lambda (p_\chi)\rangle_\chi = \langle  \frac{v_\chi}{\Gamma_\chi^{b}}\rangle_\chi \approx \frac{ \langle  v_\chi \rangle_\chi}{\langle \Gamma_\chi^{b} \rangle_\chi} = \frac{n_\chi}{\mathcal C_{\rm prod}^b} \langle  v_\chi \rangle_\chi\, . 
\end{align}

\section{Cross sections for annihilation ($2\to 2$)}
 \label{app:2to2}
 
In this appendix, we list cross sections for the $2\to 2$ processes $\ell\ell\to\chi\chi, \chi\chi\to\ell\ell$ ($s$-channel) and $\ell\chi\to\ell\chi$ ($t$-channel), where $\ell$ generically refer to any lepton, including neutrinos. 
For the effective interactions in Eq.~\eqref{eq:EFTLag}, a linear independent basis is given by the operators $O_{SS}$, $O_{PP}$, $O_{SP}$, $O_{PS}$, $O_{VV}$, $O_{AA}$, $O_{AV}$, $O_{VA}$, $O_{TT}$, $O_{T^{\prime}T}$ and we find\footnote{We disagree with Ref.~\cite{Guha:2018mli} on the $t$-channel recovering their results only in the limit $m_\ell, m_\chi\to 0$.}
\begin{widetext}

\begin{align}
\begin{split}
\sigma_{\ell\ell\to\chi\chi} =& \frac{\sqrt{s-4 m_{\chi }^2}}{48 \pi  s \Lambda_l^4 \sqrt{s-4 m_\ell^2}} \bigg[3\, C_{SS}^2(s-4m_{\ell}^2)(s-4m_{\chi}^2) + 3\, C_{PP}^2s^2 + 3\, C_{PS}^2s(s-4m_{\chi}^2) + 3\, C_{SP}^2s(s-4m_{\ell}^2)\\
& + 4\, C_{VV}^2(s+2m_{\ell}^2)(s+2m_{\chi}^2) + 4\, C_{AA}^2(s^2-4s(m_{\ell}^2+m_{\chi}^2)+28m_{\ell}^2m_{\chi}^2) + 4\, C_{VA}^2(s+2m_{\ell}^2)(s-4m_{\chi}^2)\\
&+ 4\, C_{VA}^2(s-4m_{\ell}^2)(s+2m_{\chi}^2) + 8\, C_{TT}^2(s^2+2s(m_{\ell}^2+m_{\chi}^2)+40m_{\ell}^2m_{\chi}^2) + 8\, C_{T^{\prime}T}^2(s^2+2s(m_{\ell}^2+m_{\chi}^2)-32m_{\ell}^2m_{\chi}^2)\\
& - 24\, C_{AA}\, C_{PP}\,s\,m_{\ell}m_{\chi} + 144\, C_{VV}\, C_{TT}\,s\,m_{\ell}m_{\chi}\bigg]\,,
\end{split}
\label{eq:sschannel}
\end{align}

\begin{align}
\begin{split}
\sigma_{\chi\ell\to\chi\ell} =& \frac{1}{48\pi \Lambda_l^4 s^3} \bigg [ C_{SS}^2 \Big( s^4 + 2s^3(m_\ell^2+m_\chi^2)-2s^2 (3m_\ell^4 -14m_\ell^2 m_\chi^2 +3m_\chi^4)+2s (m_\ell^2 +m_\chi^2)(m_\ell^2-m_\chi^2)^2+(m_\ell^2-m_\chi^2)^4\Big)\\
&+C_{SP}^2 \Big( m_\ell^4 -2m_\ell^2 (m_\chi^2-2s) +(s-m_\chi^2)^2\Big)\Big( m_\ell^4 -2m_\ell^2 (m_\chi^2 +s) +(s-m_\chi^2)^2\Big)\\
&+C_{PS}^2 \Big(m_\ell^4 -2m_\ell^2 (s+m_\chi^2) +(s-m_\chi^2)^2\Big)\Big(m_\ell^4 +m_\chi^4 -2m_\ell^2 (m_\chi^2 +s) +s(s+4m_\chi^2)\Big)\\
&+C_{PP}^2 \Big( m_\ell^2 -2m_\ell^2 (s+m_\chi^2) +(s-m_\chi^2)^2 \Big)^2\\
&+2C_{VV}^2 \Big( 4s^4 -10s^3 (m_\ell^2 +m_\chi^2) +s^2 (9m_\ell^2 +22m_\ell^2 m_\chi^2 +9m_\chi^4\Big)-4s(m_\ell^2 +m_\chi^2)(m_\ell^2-m_\chi^2)^2+(m_\ell^2-m_\chi^2)^4\Big)\\
&+2\, C_{VA}^2 \Big( s-(m_\ell+m_\chi)^2\Big)\Big( s-(m_\ell-m_\chi)^2\Big)\Big( m_\ell^4 -2m_\ell^2 (m_\chi^2 +s) +(2s+m_\chi^2)^2\Big)\\
&+2\, C_{AV}^2 \Big( s-(m_\ell+m_\chi)^2\Big)\Big( s-(m_\ell-m_\chi)^2\Big)\Big( 4s^2+2s(2m_\ell^2-m_\chi^2) +(m_\ell^2-m_\chi^2)^2\Big)\\
&+2\, C_{AA}^2 \Big( 4s^4 -4s^3(m_\ell^2+m_\chi^2) -s^2(3m_\ell^4 -46m_\ell^2m_\chi^2 +3m_\chi^4)+2s(m_\ell^2+m_\chi^2)(m_\ell^2-m_\chi^2)^2 +(m_\ell^2-m_\chi^2)^4\Big)\\
& + 8\, C_{TT}^2\Big( 7s^4 -13s^3 (m_\ell^2+m_\chi^2) +2s^2 (3m_\ell^4+26m_\ell^2m_\chi^2 +3m_\chi^4)-s(m_\ell^2+m_\chi^2)(m_\ell^2-m_\chi^2)^2 + (m_\ell^2-m_\chi^2)^4\Big)\\
&+8\, C_{T^{\prime}T}^2 \Big( s-(m_\ell+m_\chi)^2\Big)\Big( s-(m_\ell-m_\chi)^2\Big)\Big( 7s^2 +s(m_\ell^2+m_\chi^2) +(m_\ell^2-m_\chi^2)^2\Big)\\
&+4\, C_{SS}\Big(C_{TT} (m_{\ell}^8-m_{\ell}^6 (4 m_{\chi}^2+s)+m_{\ell}^4 m_{\chi}^2 (6 m_{\chi}^2+s)+m_{\ell}^2 (-4 m_{\chi}^6+m_{\chi}^4 s-8 m_{\chi}^2 s^2+5 s^3)\\
&+m_{\chi}^8-m_{\chi}^6 s+5 m_{\chi}^2 s^3-2 s^4)-3\, C_{VV} m_{\ell} m_{\chi} s (m_{\ell}^4-2 m_{\ell}^2 (m_{\chi}^2-s)+m_{\chi}^4+2 m_{\chi}^2 s-3 s^2)\Big)\\
& + 4\, C_{PP} \Big(3\, C_{AA}  m_{\chi} s (m_{\ell}^4-2m_{\ell}^2 (m_{\chi}^2+s)+(m_{\chi}^2-s)^2)+ C_{TT} (m_{\ell}^8-m_{\ell}^6 (4 m_{\chi}^2+s)+m_{\ell}^4 m_{\chi}^2 (6 m_{\chi}^2+s)\\
& +m_{\ell}^2 (-4 m_{\chi}^6+m_{\chi}^4 s-8 m_{\chi}^2 s^2+5 s^3)+m_{\chi}^8-m_{\chi}^6 s+5 m_{\chi}^2 s^3-2 s^4)\Big)\\
&+ 4(C_{SP}+C_{PS})\Big(C_{T^{\prime}T} (m_{\ell}^8-m_{\ell}^6 (4 m_{\chi}^2+s)+m_{\ell}^4 m_{\chi}^2 (6 m_{\chi}^2+s)+m_{\ell}^2 (-4 m_{\chi}^6+m_{\chi}^4 s-8 m_{\chi}^2 s^2+5 s^3)+m_{\chi}^8\\
&-m_{\chi}^6 s+5 m_{\chi}^2 s^3-2 s^4)+6 C_{VA} m_{\ell} m_{\chi} s^2 (m_{\chi}^2-m_{\ell}^2)\Big)\\
& -4\, C_{VV}\Big(C_{AA} (m_{\ell}^8-m_{\ell}^6 (4 m_{\chi}^2+s)+m_{\ell}^4 m_{\chi}^2 \left(6 m_{\chi}^2+s\right)+m_{\ell}^2 (-4 m_{\chi}^6+m_{\chi}^4 s-8 m_{\chi}^2 s^2+5 s^3)+m_{\chi}^8\\
&-m_{\chi}^6 s+5 m_{\chi}^2 s^3-2 s^4)+18 C_{TT} m_{\ell} m_{\chi} s (m_{\ell}^4-2 m_{\ell}^2 (m_{\chi}^2+s)+(m_{\chi}^2-s)^2)\Big)\\
&+72\, C_{AA}\, C_{TT} m_{\ell} m_{\chi} s(m_{\ell}^4-2 m_{\ell}^2 \left(m_{\chi}^2-s\right)+m_{\chi}^4+2 m_{\chi}^2 s-3 s^2)\\
&+ C_{VA}\Big( 144C_{T^{\prime}T} m_{\ell} m_{\chi} s^2 (m_{\ell}^2-m_{\chi}^2)-4 C_{AV} (m_{\ell}^8-m_{\ell}^6 (4 m_{\chi}^2+s)+m_{\ell}^4 m_{\chi}^2 (6 m_{\chi}^2+s)\\
&+m_{\ell}^2 (-4 m_{\chi}^6+m_{\chi}^4 s-8 m_{\chi}^2 s^2+5 s^3)+m_{\chi}^8-m_{\chi}^6 s+5 m_{\chi}^2 s^3-2 s^4)\Big)\\
&+144\, C_{AV}\, C_{T^{\prime}T}\, s^2m_{\ell}m_{\chi}(m_{\ell}^2-m_{\chi}^2) \bigg]\,.
\end{split}
\label{eq:stchannel}
\end{align}

\end{widetext}

We can also generalize the cross sections for the $Z^\prime$ model in Eq.~\eqref{eq:LagSimplified} by including generic vector ($V_i$) and axial ($A_i$) couplings to the leptons (such that $V_\ell=1$, $A_\ell=0$ gives back the model in Eq.~\eqref{eq:LagSimplified}):
\begin{align}
\sigma_{\ell\ell\to\chi\chi}^V =& \frac{g_\ell^2 g_\chi^2}{12\pi s} \sqrt{\frac{s-4m_\chi^2}{s-4m_\ell^2}} \frac{s+2m_\chi^2}{(s-m_{Z'}^2)^2 + m_{Z'}^2 \Gamma_{Z'}^2} \nonumber\\
&\times\Big[ V_\ell^2 (s+2m_\ell^2) + A_\ell^2 (s-4m_\ell^2)\Big]\, .\nonumber
\label{eq:xs2}
\end{align}
The $t$-channel for small $Z'$ masses is more involved because the propagator depends on the Mandelstam variable $t$, over which one integrates to obtain the total cross section. Neglecting the $Z'$ decay width, the resulting expression reads
\begin{align}
\sigma_{\chi\ell\to\chi\ell}^V&= \frac{g_\ell^2 g_\chi^2}{8\pi} \Big[ (V_\ell^2 +A_\ell^2) \frac{2s+m_{Z'}^2}{sm_{Z'}^2}\\
&+ \frac{1}{m_{Z'}^2} \frac{V_\ell^2 (m_{Z'}^4 +8m_\ell^2 m_\chi^2) + A_\ell^2 m_{Z'}^2 (m_{Z'}^2-4m_\ell^2)}{m_\ell^4 -2m_\ell^2 (s+m_\chi^2) +(s-m_\chi^2)^2 +sm_{Z'}^2} \nonumber\\
&-2\frac{V_\ell^2 (s+m_{Z'}^2) +A_\ell^2 (s + m_{Z'}^2 - 2m_\ell^2)}{m_\ell^4 -2m_\ell^2 (s+m_\chi^2) +(s-m_\chi^2)^2}\nonumber\\
&\times\log \frac{m_\ell^4 -2m_\ell^2 (s+m_\chi^2) +(s-m_\chi^2)^2 +sm_{Z'}^2}{sm_{Z'}^2} \Big]\, .\nonumber
\label{eq:xs5}
\end{align}

The cross sections for the inverse process $\chi\chi\to\ell\ell$ can be obtained using
\begin{equation}
\sigma_{\chi\chi\to\ell\ell} = \frac{s-4m_\ell^2}{s-4m_\chi^2} \sigma_{\ell\ell\to\chi\chi}\, .
\label{eq:xs3}
\end{equation}

\section{Cross sections for photoproduction ($2\to 3$) in the EFT limit}
\label{app:2to3}

In this appendix, we derive the cross sections for the photoproduction processes $\ell^-\gamma \to \ell^- \bar\chi \chi$ with the effective operators given in Eq.~\eqref{eq:EFTLag}. For simplicity, we rewrite Eq.~\eqref{eq:EFTLag}  factorizing the Wilson Coefficients in terms of a leptonic and dark current, $C_{XY}^l=X_lY_\chi$. 
We start with the most simple case of scalar interactions. The amplitude reads 
\begin{align}
i\mathcal M =& \frac{e}{\Lambda_\ell^2} \epsilon_\mu (p_b) \bar u_\chi (p_1) (S_\chi + iP_\chi \gamma_5) v_\chi (p_2) \nonumber\\
& \times \bar u_\ell (p_3) \Big[ (S_\ell + iP_\ell \gamma_5 ) \frac{\slashed{p}_a + \slashed{p}_b +m_\ell }{(p_a+p_b)^2 - m_\ell^2} \gamma^\mu \nonumber\\
& + \gamma^\mu \frac{\slashed{p}_3 - \slashed{p}_b + m_\ell}{(p_3 - p_b)^2 - m_\ell^2} (S_\ell + iP_\ell \gamma_5)\Big] u_\ell (p_a)\, . 
\end{align}
The squared and spin-averaged amplitude factorizes in two contributions
\begin{align}
\overline{|\mathcal{M}|^2} =  X (p_1, p_2) L (p_a, p_b, p_3)\, ,
\end{align}
where the $X$ and $L$ denote the traces over dark and SM particles, respectively. The phase space can be factorized
\begin{align}
d\Phi_3 &(p_a + p_b; p_1, p_2, p_3)=\nonumber\\ &\frac{dm_{12}^2}{2\pi}d\Phi_2 ( p_a + p_b; p_{12}, p_3) d\Phi_2 ( p_{12}; p_1,p_2)\, ,
\label{phi3}
\end{align}
where we introduced $s=(p_a+p_b)^2$ and the momentum of the effective two-body system of dark particles $p_{12} = p_1 + p_2$ with invariant mass $m_{12}^2 = p_{12}^2$. 

We start with the dark system, i.e. $d\Phi_2 ( p_{12}; p_1,p_2)$. The function $X (p_1, p_2)$ is a scalar and can therefore only depend on the scalar product $p_1 p_2$, which can be rewritten in terms of $m_{12}^2$ using $p_1 p_2 = (m_{12}^2 - 2m_\chi^2)/2$, leading to $X (p_1, p_2) = \tilde X(m_{12})$. Thus, we obtain
\begin{align}
\int d\Phi_2 (p_{12}; p_1, p_2) X(p_1, p_2)= \frac{\tilde X(m_{12})}{8\pi} \sqrt{1-4\frac{m_\chi^2}{m_{12}^2}}\, .\nonumber
\end{align}
The second phase-space integral can be simplified as
\begin{align}
	\int d\Phi_2 (p_a+p_b;& p_{12}, p_3) = \frac{1}{(4\pi)^2\sqrt{s}}\nonumber\\
 &\times\int \bar p_3 d\bar p_3 \delta \Big(\bar p_3 - \frac{\sqrt{s}}{2} \beta \Big) d\cos\theta d\phi\, , 
 \end{align}
 with
\begin{align}
\beta &= \sqrt{1- 2\frac{m_{12}^2 +m_\ell^2}{s} + \frac{(m_{12}^2 - m_\ell^2)^2}{s^2}}\, ,
\label{beta}
\end{align}
where $\bar p_3$ is the spatial component of the 4-vector $p_3$, i.e. $p_3^2 = E_3^2 -\bar p_3^2 = m_\ell^2$. 

The $p_3$-dependence in the function $L(p_a, p_b, p_3)$ can be rewritten in terms of $s, \bar p_3$ and $\cos\theta$ in the center-of-mass (CM) frame using
\begin{align}
\begin{split}
p_a p_3 &= \frac{\sqrt{s}}{2} \Big[ E_3 \Big( 1+\frac{m_\ell^2}{s}\Big) - \bar p_3 \cos\theta \Big( 1-\frac{m_\ell^2}{s}\Big)\Big]\, , \\
p_b p_3 &= \frac{\sqrt{s}}{2} \Big( 1-\frac{m_\ell^2}{s}\Big) (E_3 + \bar p_3 \cos\theta)\, .
\end{split}
\end{align}
One can now evaluate the $d\Phi_2 (p_a+p_b; p_{12},p_3)$ integral. After the trivial integrals in $d\phi$ and $d\bar p_3$ one can perform the $d\cos\theta$ integration analytically. At this point, we are only left with the $m_{12}^2$ integration, which can not be done analytically in the general case. Introducing the dimensionless variable
\begin{align}
\beta_\chi=\sqrt{1-4m_\chi^2/m_{12}^2}\, , 
\end{align}
and $x = m_{12}^2 / m_\ell^2, x_\chi = m_\chi^2 / m_\ell^2, \hat s = s/m_\ell^2$, we arrive at the result
\begin{align}
\begin{split}
\sigma =& \frac{\alpha m_\ell^2}{64\pi^2\Lambda_\ell^4} \frac{1}{(\hat s-1)^3 \hat s^2} \int_{4x_\chi}^{(\sqrt{\hat s}-1)^2} dx\,x\,\beta_\chi \\
&\times \Big(\beta_\chi^2S_\chi^2  + P_\chi^2 \Big)\Big[ S_\ell^2 f_S (x, \hat s) + P_\ell^2 f_P (x,\hat s)\Big]\, ,
\end{split}
\end{align}
where the functions $f_S, f_P$ are defined by
\begin{align}
f_S =& 4\hat s^2 \Big( \hat s^2 -2\hat s(x-3) + 2x^2 -10x+9\Big) T(x,\hat s),\nonumber\\
&-\Big( 3\hat s^3 +\hat s^2 (25-7x)  +\hat s (5-2x) +x-1\Big) R(x,\hat s)\, ,\nonumber\\
f_P =& 4\hat s^2 \Big( \hat s^2 -2\hat s(x+1) + 2x^2 -2x+1\Big) T(x,\hat s),\nonumber\\
&-\Big( 3\hat s^3 - 7\hat s^2 (x+1) +\hat s(5-2x) +x -1\Big) R(x,\hat s)\, ,
\end{align}
with,
\begin{align}
R(x,\hat s) =& \sqrt{\hat s^2 -2\hat s (x+1) +(x-1)^2} \, ,\nonumber\\
T(x,\hat s) =& \tanh^{-1} \frac{R(x,\hat s)}{\hat s-x+1}\, .
\end{align}
In the limit $m_\chi\to 0$, the final integral can be evaluated analytically,
\begin{align}
\sigma&=\frac{\alpha m_\ell^2}{2304\pi^2 \Lambda_\ell^4} \frac{1}{(\hat s-1)^3 \hat s^2}\nonumber\\
&\times\Big( S_\chi^2 + P_\chi^2\Big) \Big( S_\ell^2 g_S (\hat s) + P_\ell^2 g_P (\hat s)\Big)\, ,
\end{align}
where
\begin{align}
\begin{split}
g_S =& -79\hat s^6 -14 \hat s^5 -189\hat s^4 -296\hat s^3 +527\hat s^2 +54\hat s -3 \\
&+ 12 \hat s^2 \Big( 2\hat s^4 +10 \hat s^3 +9\hat s^2 +44\hat s+25\Big) \log \hat s \, ,\\
g_P =&- 79\hat s^6 +338\hat s^5 +675\hat s^4 -1160\hat s^3 +175\hat s^2 +54\hat s \\
&-3+12 \hat s^2 \Big( 2\hat s^4 +2\hat s^3 -63\hat s^2 -28\hat s+17\Big) \log \hat s\, .
\end{split}
\end{align}

The matrix element for vector-like interactions reads
\begin{align}
i\mathcal{M} =& \frac{e}{\Lambda_\ell^2} \epsilon_\mu (p_b) \bar u_\chi (p_1) \gamma_\nu (V_\chi + iA_\chi \gamma_5) v_\chi (p_2) \\
& \times \bar u_\ell (p_3) \Big[ \gamma^\nu (V_\ell + i A_\ell \gamma_5 ) \frac{\slashed{p}_a + \slashed{p}_b +m_\ell }{(p_a+p_b)^2 - m_\ell^2} \gamma^\mu \nonumber\\
&+ \gamma^\mu \frac{\slashed{p}_3 - \slashed{p}_b + m_\ell}{(p_3 - p_b)^2 - m_\ell^2} \gamma^\nu(V_\ell + iA_\ell \gamma_5)\Big] u_\ell (p_a)\, . \notag
\end{align}
After squaring and spin-averaging, one again obtains two traces for the SM and dark part of the amplitude, respectively. Due to the vector-like nature of the interaction, these traces are contracted with two Lorentz indices, one arising from $\mathcal{M}$ and one from $\mathcal{M}^\dagger$
\begin{align}
\overline{|\mathcal{M}|^2}  =  X_{\mu\nu} (p_1, p_2) L^{\mu\nu} (p_a, p_b, p_3)\, .
\label{photoproductionV}
\end{align}
Due to the fact that $X_{\mu\nu}$ can not only depend on $m_{12}^2$, but also on $p_1^\mu, p_2^\mu$, we can not simply factor $X_{\mu\nu}$ out of the $d\Phi (p_{12}; p_1, p_2)$ integral as in the scalar case. However, Lorentz covariance implies that the integral can only depend on the vector $p_{12}^\mu =p_1^\mu +p_2^\mu$
\begin{align}
\int d\Phi_2 (p_{12}; p_1, p_2) X^{\mu\nu} (p_1, p_2) = A_1 m_{12}^2  g^{\mu\nu} + A_2 p_{12}^\mu p_{12}^\nu\, ,
\label{photoproductionV2}
\end{align}
where $A_1, A_2$ are functions of $m_{12}^2$ 
\begin{align}
A_1 &= -\Big( (V_\chi^2 + A_\chi^2) + 2\frac{m_\chi^2}{m_{12}^2} (V_\chi^2 - 2A_\chi^2)\Big)\, , \nonumber\\
A_2 &=\left(1+\frac{2m_\chi^2}{m_{12}^2}\right) (V_\chi^2 + A_\chi^2)\, .
\end{align}
From this point on, the calculation is equivalent to the scalar case. We arrive at the result
\begin{align}
\sigma =& \frac{\alpha m_\ell^2}{48\pi^2 \Lambda_\ell^4}\frac{1}{(\hat s-1)^3 \hat s^2}\\
&\times \int_{4x_\chi}^{(\sqrt{\hat s}-1)^2} dx\,\beta_\chi \Big( V_\ell^2 f_V (x, x_\chi, \hat s) + A_\ell^2 f_A (x, x_\chi, \hat s)\Big)\, ,\nonumber
\end{align}
with
\begin{align}
f_V&=-4A_1 \hat s^2 x\Big( \hat s^2-2\hat s (x+3) +2x^2 +2x-3\Big) T(x,\hat s)\nonumber\\
&-A_1 x \Big( \hat s^3+\hat s^2 (7x+15) +\hat s (2x-1) -x+1\Big) R(x,\hat s)\, , \nonumber\\
f_A&= -4A_1 \hat s^2 x\Big( \hat s^2 -2\hat s(x-5) +2x^2 -14x+13\Big)T(x,\hat s)\nonumber\\
&-A_1 x\Big( \hat s^3+7\hat s^2 (x-7) + \hat s(2x-1)-x+1\Big) R(x,\hat s)\nonumber\\
&+8 A_2 \hat s^2 \Big( \hat s^2 -2\hat s(x+1) +2x^2 -2x+1\Big) T(x,\hat s)\nonumber\\
&-2A_2 \Big( 3\hat s^3 -7\hat s^2 (1+x) +\hat s(5-2x)+x-1\Big) R(x,\hat s)\, .
\end{align}

In the limit $m_\chi\to 0$, one can again perform the $dx$ integration analytically and arrive at
\begin{align}
\sigma &= \frac{\alpha m_\ell^2}{1728 \pi^2 \Lambda_\ell^4} \frac{1}{\hat s^2 (\hat s-1)^3}\nonumber\\
& \times (V_\chi^2 + A_\chi^2) \Big( V_\ell^2 g_V (\hat s) + A_\ell^2 g_A (\hat s)\Big)\, ,
\end{align}
with
\begin{align}
g_V =& -55\hat s^6+682\hat s^5 + 483\hat s^4 -968\hat s^3 -169\hat s^2 +30\hat s \nonumber\\
&-3+ 12\hat s^2\Big( 2\hat s^4 -14\hat s^3 -87\hat s^2 -52\hat s+1\Big) \log\hat s \, ,\nonumber\\
g_A =& -55\hat s^6-254\hat s^5 +219\hat s^4-296\hat s^3 +119\hat s^2 +294\hat s\nonumber\\
&-27+12\hat s^2 \Big( 2\hat s^4+10\hat s^3 +33\hat s^2-4\hat s + 49\Big)\log\hat s\, .\nonumber
\end{align}

Finally, the matrix element for tensor-like interactions reads
\begin{align}
i\mathcal{M} =& \frac{e}{\Lambda_\ell^2} \epsilon_\mu (p_b) \bar u_\chi (p_1) \sigma_{\nu\rho} (T_\chi +  T'_\chi \gamma_5) v_\chi (p_2) \\
& \times \bar u_\ell (p_3) \Big[ \sigma^{\nu\rho} \frac{\slashed{p}_a + \slashed{p}_b +m_\ell }{(p_a+p_b)^2 - m_\ell^2} \gamma^\mu \\
& + \gamma^\mu \frac{\slashed{p}_3 - \slashed{p}_b + m_\ell}{(p_3 - p_b)^2 - m_\ell^2} \sigma^{\nu\rho}\Big] u_\ell (p_a)\, . 
\end{align}
The squared and spin-averaged matrix element can be factorized as
\begin{align}
\overline{|\mathcal{M}|^2}  = X_{\mu\nu\rho\sigma} (p_1, p_2) L^{\mu\nu\rho\sigma} (p_a, p_b, p_3)\, .
\end{align}
We define the index ordering in $X^{\mu\nu\rho\sigma}$ such that the first and second pair of indices each corresponds to one $\sigma^{\mu\nu}$ in the trace that constitutes $X^{\mu\nu\rho\sigma}$, i.e. $X^{\mu\nu\rho\sigma}$ is antisymmetric under $\mu\leftrightarrow \nu$ and $\rho\leftrightarrow\sigma$. After proper antisymmetrization, there are only two Lorentz structures that satisfy this condition, leading to
\begin{align}
\int &d\Phi_2 (p_{12}; p_1, p_2) X^{\mu\nu\rho\sigma} (p_1, p_2) = m_{12}^2  B_1 \Big( g^{\mu\rho}g^{\nu\sigma} - g^{\nu\rho} g^{\mu\sigma}\Big)\nonumber\\
&+ B_2 \Big( p_{12}^\mu (p_{12}^\rho g^{\nu\sigma} - p_{12}^\sigma g^{\nu\rho}) - p_{12}^\nu (p_{12}^\rho g^{\mu\sigma} - p_{12}^\sigma g^{\mu\rho})\Big)\, ,\nonumber
\end{align}
where $B_1, B_2$ are functions of $m_{12}^2$
\begin{align}
B_1 =& \frac{1}{12\pi}\beta_\chi \Big( (T_\chi^2 + T'^2_\chi) + 4\frac{m_\chi^2}{m_{12}^2} (2T'^2_\chi - T_\chi^2)\Big)\, ,\nonumber \\
B_2 =& -\frac{1}{6\pi} \beta_\chi \frac{m_{12}^2 + 2m_\chi^2}{m_{12}^2} (T_\chi^2 + T'^2_\chi)\, .\nonumber   
\end{align}
After following the standard procedure as outlined for the scalar case, one arrives at the result
\begin{align}
\sigma =& \frac{\alpha m_\ell^2}{4\pi \Lambda_\ell^4} \frac{1}{(\hat s-1)^3 \hat s^2}\\
&\int_{4x_\chi}^{(\sqrt{\hat s}-1)^2} dx \Big(B_1 f_1 (x, x_\chi, \hat s) + B_2 f_2 (x, x_\chi, \hat s)\Big)\, ,\nonumber
\end{align}
where
\begin{align}
f_1 =& 96\hat s^2\Big( (\hat s-x+1) T(x,\hat s)-  R(x,\hat s)\Big)\, , \nonumber\\
f_2 =& -\Big( 4\hat s^4 + \hat s^3 (x-16) + \hat s^2 (7x^2 + 59 x+24) \nonumber\\
&+ \hat s (2x^2 + 7x-16) -x^2 -3x+4\Big)R(x,\hat s) \nonumber\\
&-4\hat s^2 x\Big( \hat s^2 -2\hat s (x+9) +2x^2 + 14x - 15\Big)  T(x,\hat s)\, .
\end{align}
In the limit $m_\chi\to 0$, the final $dx$ integration can be done analytically and we arrive at
\begin{align}
\sigma =& \frac{\alpha m_\ell^2}{864\pi^2 \Lambda_\ell^4} \frac{1}{(\hat s-1)^3 \hat s^2} (T_\chi^2 + T'^2_\chi) \\
&\times\Big[ 17\hat s^6+370\hat s^5+675\hat s^4-344\hat s^3-1153\hat s^2+486\hat s -51\nonumber\\
&+12\hat s^2 \Big( 2\hat s^4-26\hat s^3 -27\hat s^2 -136\hat s+37\Big) \log\hat s \Big]\, .\nonumber
\end{align}

\section{Full calculation of photoproduction}
\label{app:full_photoproduction}
The energy-loss rate per unit volume for photoproduction $\ell^{-}  (p_a)\gamma (p_b) \to \chi (p_1) \overline{\chi} (p_2) \ell^{-}  (p_3) $ can be written as
\begin{align}
Q &  = \frac{1}{32 \pi^4} \int_{m_\ell}^\infty d E_a  \bar{p}_a f_a  \int_0^\infty d E_b   E_b  f_b  \nonumber \\ & \times  \int_{-1}^1 d c_\theta J_s (s, E_a ,E_b) \, ,
\end{align}
\begin{align}
J_s & = \int d \Phi_3 (p_a + p_b ; p_1, p_2, p_3) |{\cal M}|^2   \nonumber \\
& \times  (E_a + E_b - E_3)  (1-f_3) \, , 
\label{Js}
\end{align}
where $d \Phi_3$ is the Lorentz-invariant 3-body phase space volume,  $\bar{p}_a \equiv |\vec{p}_a| = \sqrt{E_a^2 - m_\ell^2}$, $f_a^{-1}= e^{(E_a - \mu_\ell)/T}+1$, $f_b^{-1}= e^{E_b/T} - 1$, $f_3^{-1}= e^{(E_3 - \mu_\ell)/T}+1$, $s = m_\ell^2 + 2 E_b (E_a - \bar{p}_a c_\theta)$, $c_\theta \equiv \cos \theta$ and all energies refer to the PNS frame.

Using the decomposition of the 3-body phase space in Eq.~\eqref{phi3}, the 2-body phase space integral over the dark fermion momenta $p_1$ and $p_2$ can be easily performed, as the spin-summed, squared matrix element $|{\cal M}|^2$ factorizes into two contributions, cf. Eq.~\eqref{photoproductionV}
\begin{align}
M (p_a, p_b, p_3, m_{12}) & \equiv \int d \Phi_2 (p_{12}; p_1, p_2) \left|{\cal M} \right|^2 \, .
\label{M}
 \end{align}
The remaining phase space integral $d \Phi_2 (p_a + p_b ; p_{12}, p_3)$ is suitably evaluated in the CM frame, giving 
\begin{align}
 \int d \Phi_2   & = \frac{2 \pi}{\sqrt{s}} \int \frac{\bar{p}_3^\prime d \bar{p}_3^\prime d c_\delta d \phi}{4 (2 \pi)^3} \delta (\bar{p}_3^\prime - \beta \sqrt{s}/2) \, , 
 \label{phi2}
\end{align}
where $c_\delta \equiv \cos \delta$, $\bar{p}_3^\prime = \sqrt{(E_3^\prime)^2 - m_\ell^2}$ with  $E_3^\prime$ the final lepton energy in the CM frame, $\phi$ and $\delta$ denote the polar and azimuthal angle of ${\vec{p}_3}^{\, \prime}$ and $\beta (s, m_{12})$ is given in Eq.~\eqref{beta}.

To evaluate the Lorentz-invariant quantity $M (p_a, p_b, p_3, m_{12}^2)$ in this frame, we express  the relevant Lorentz scalars in the CM frame, where $p_a^\prime = (E_a^\prime, 0, 0, E_b^\prime)$, $p_b^\prime = (E_b^\prime, 0, 0, - E_b^\prime)$ 
\begin{align}
p_a p_b & = \frac{s-m_\ell^2}{2} \, , \nonumber \\ p_a p_3 & = E_a^\prime E_3^\prime + E_b^\prime \bar{p}_3^\prime c_\delta \,  , \nonumber \\  p_b p_3 & = E_b^\prime E_3^\prime - E_b^\prime \bar{p}_3^\prime c_\delta \, , 
\label{prel}
\end{align}
where the energies in the CM frame can be written in terms of $s$ only using
\begin{align}
E_a^\prime & =\frac{s + m_\ell^2}{2 \sqrt{s}}  \, , & E_b^\prime & = \frac{s-m_\ell^2}{2 \sqrt{s}} \, .
\label{Eabprime}
\end{align}
These expressions allow to write $M (p_a, p_b, p_3, m_{12})$ as a function of $s, \cos \delta, E_3^\prime$ and $m_{12}$, and it remains to evaluate $(E_a + E_b - E_3) (1-f_3)$ in the CM frame. $E_3$ is related to $E_3^\prime$ by a Lorentz boost with velocity $(\vec{p}_a + \vec{p}_b)/(E_a + E_b)  \equiv \vec{p}_{ab}/E_{ab} = \bar{p}_{ab}/E_{ab} (0, s_\eta,  c_\eta)$ with $s_\eta \equiv \sin \eta$, $c_\eta \equiv \cos \eta$ and boost factor $\gamma = E_{ab}/\sqrt{s}$ (since $s = E_{ab}^2 - \bar{p}_{ab}^2$). The same boost relates $E_b$ to $E_b^\prime$, which can be used to obtain an expression for $\cos \eta$ as 
a function of $s$ and the initial energies in the PNS frame
\begin{align}
c_\eta & = \frac{s (E_b-E_a)+ m_\ell^2 (E_b + E_a) }{(s-m_\ell^2) \sqrt{(E_a + E_b)^2-s} } \, .
\label{eta}
\end{align}
This angle determines the desired relation between $E_3$ and $E_3^\prime$ as
\begin{align}
E_3 & = \frac{1}{\sqrt s}  \left( E_{ab} E_3^\prime +  \bar{p}_{ab}  \bar{p}_3^{\, \prime}   (s_\eta s_\phi s_\delta + c_\eta c_\delta ) \right)\, .
\label{E3}
\end{align}
Having expressed all final state energies in the CM frame, one can finally perform the $d \bar{p}_3^\prime$ using the $\delta$-function in Eq.~\eqref{phi2}, giving
\begin{align}
\bar{p}_3^\prime & = \beta \frac{\sqrt s}{2} \, , & E_3^\prime & = \frac{\sqrt s}{2} \left( 1 - \frac{m_{12}^2 - m_\ell^2}{s} \right) \, .
\label{E3prime}
\end{align}
Putting everything together, we finally obtain for Eq.~\eqref{Js}
\begin{align}
J_s & = \frac{1}{64 \pi^3} \int_{4 m_\chi^2}^{(\sqrt{s} - m_l)^2} dm^2_{12}\,  \beta (s, m_{12})  \int_{-1}^1 d c_\delta M(s,m_{12}^2, c_\delta) \nonumber \\ & \times \int_0^{2 \pi} d\phi (E_a + E_b - E_3) (1-f_3) \, , 
\end{align}
 where $\beta (s, m_{12})$ is given in Eq.~\eqref{beta},  $M(s,m_{12}^2, c_\delta)$ is obtained from Eqs.~\eqref{M},\eqref{prel},\eqref{Eabprime},\eqref{E3prime}, and $E_3 = E_3 (s, m_{12}, E_a, E_b, \phi, \delta )$ from Eqs.~\eqref{eta},\eqref{E3},\eqref{E3prime}.
When $1-f_3$ is to good approximation independent of $E_3$, the $d \phi$ integration can be trivially carried out, giving an overall factor of $2 \pi$ and $ \left(s_\eta s_\phi s_\delta + c_\eta c_\delta  \right) \to  c_\eta c_\delta $ in Eq~\eqref{E3}.

\section{Numerical resolution of the kinematic integrals for the annihilation rates}
\label{app:numerical}
Throughout this work, we use Monte Carlo techniques~\cite{Lepage:2020tgj} to evaluate the thermal phase space integrals numerically, e.g. Eq.~\eqref{eq:Qann} for annihilation and Eq.~\eqref{eq:GammaAnn} for trapping. For processes where the $Z'$ can be produced resonantly, the Breit-Wigner propagator yields a peak in the Mandelstam variable $s$ located at $m_{Z'}$ with width $\sqrt{m_{Z'}\Gamma_{Z'}}\sim g m_{Z'}$. We want to calculate these integrals for very small couplings $g\sim 10^{-10}$, translating into very narrow peaks. When using only several thousands of points for the Monte Carlo estimate, it is unlikely that this estimate can resolve the peak structure. In particular, iterative algorithms like VEGAS are unable to adapt to the peak if they can not extract information about the peak from samples in early iterations.

Fortunately, we have knowledge about the shape of the peak, allowing us to provide this information to the Monte Carlo sampler. To achieve this, we include the Mandelstam variable $s$, corresponding to the peak location, as one of the integration variables. Subsequently, we divide the integral into three regions. One of these regions includes only the peak 
\begin{equation}
    m_{Z'}^2 - \sigma m_{Z'}\Gamma_{Z'}\le s \le m_{Z'}^2 + \sigma m_{Z'}\Gamma_{Z'}\, , 
\end{equation}
with $\sigma\sim 5$. The other two regions cover the regions to the left and to the right of the peak, respectively. This approach ensures that the sum of the three integrals effectively captures and resolves the peak. 

\begin{figure*}[t]

\begin{tabular}{cc} 
\includegraphics[width=1.0\columnwidth]{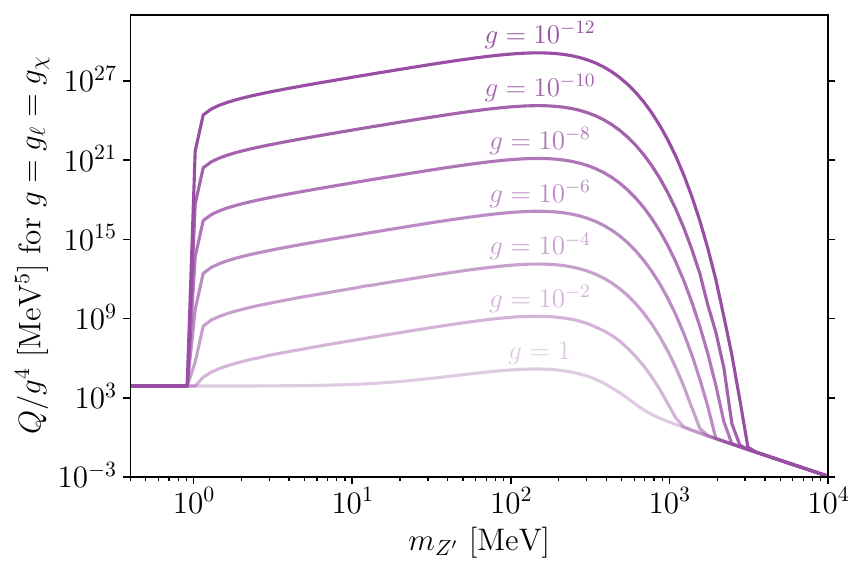} &
\includegraphics[width=1.0\columnwidth]{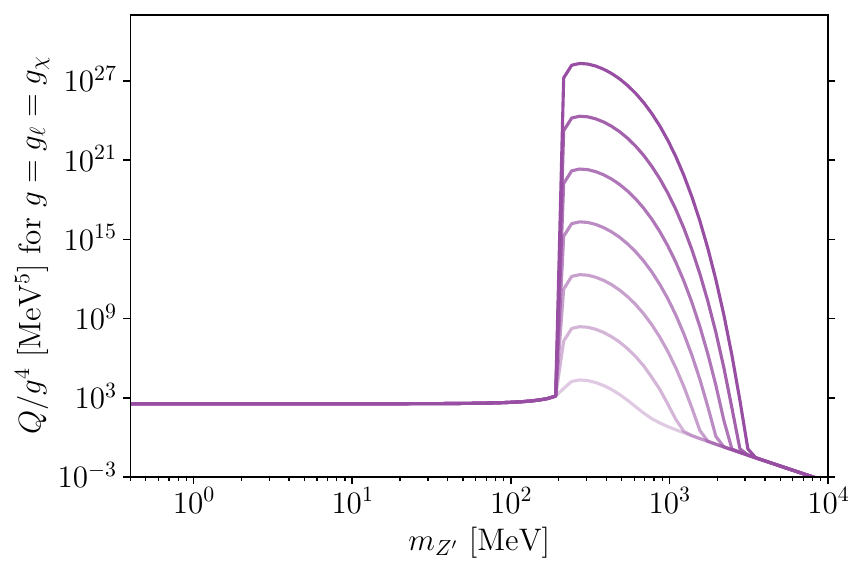}
\end{tabular}
\caption{Free-streaming luminosity at R=9.2 km for a $Z'$ coupled to electrons (left panel) and muons (right panel), rescaled by a factor $g^4$ to highlight the regions where the resonant contribution dominates.\label{fig:resonance}}

\end{figure*}

These integrals scale as $\propto g^2$ when the contribution from the resonance dominates the integral, otherwise they scale as $\propto g^4$. We can employ this observation to validate our approach of calculating the integrals, as shown in Fig.~\ref{fig:resonance}.

\bibliographystyle{JHEP}
\bibliography{muonSN}
\end{document}